\documentclass[aps,prb,superscriptaddress,twocolumn,tightenlines]{revtex4-1}
\usepackage{amsfonts}
\usepackage{amsmath}
\usepackage{amssymb}
\usepackage[final,dvips]{graphicx}
\providecommand{\U}[1]{\protect\rule{.1in}{.1in}}
\begin{document}
\title{Mesoscopic variations of local density of states in disordered superconductors}
\author{A.~E.~Koshelev$^{1}$ and A.~A.~Varlamov}
\affiliation{Materials Science Division, Argonne National Laboratory, 9700 S.Cass Avenue,
Argonne, Illinois 60637, USA}
\affiliation{CNR-SPIN, Viale del Politecnico 1, I-00133 Rome, Italy}
\date{\today }

\begin{abstract}
We explore correlations of inhomogeneous local density of states (LDoS) for
impure superconductors with different symmetries of the order parameter (s-wave
and d-wave) and different types of scatterers (elastic and magnetic
impurities). It turns out that the LDoS correlation function of superconductor
always slowly decreases with distance up to the phase-breaking length
$l_{\phi}$ and its long-range spatial behavior is determined only by the
dimensionality, as in normal metals. On the other hand, the energy dependence
of this correlation function is sensitive to symmetry of the order parameter
and nature of scatterers. Only in the simplest case of s-wave superconductor
with elastic scatterers the inhomogeneous LDoS is directly connected to the
corresponding characteristics of normal metal. We found that in presence of
pair-breaking scattering relative LDoS variations increase with decreasing
energy.
\end{abstract}
\maketitle

\section{Introduction}

Classical theory of superconductivity for impure materials deals with average
quantities. Within the BCS approach the average fundamental characteristics of
s-wave superconductor such as transition temperature, gap, and density of
states are not sensitive to potential disorder. This statement, known as
Anderson theorem, is, in fact, not rigid at all. In particular, it is enough
to introduce in s-wave superconductors some amount of magnetic impurities and
they suppress the transition temperature and gap in the quasiparticle
spectrum. Moreover, in the case of more complex symmetry of the order
parameter, even elastic impurities depress superconductivity.

It is worth to note that average parameters do not completely describe
properties of impure materials, because, in addition, disorder induces random
point-to-point variations of all quantities. For instance, it is well known
since 60's, that the LDoS of normal metal near an isolated impurity
experiences so called Friedel oscillations at the atomic
scale.\cite{SolidStTextbook}

At the end of 70's -- beginning of 80's the theory of weak localization was
developed which described the corrections to \emph{average} values of transport
characteristics of impure electron systems caused by the quantum interference
of electrons due to their multiple impurity scattering.\cite{WeakLocal} Even
though these corrections of quantum origin were found to be small in comparison
to the corresponding classical values, it was demonstrated that they have
nontrivial dependences on temperature, frequency, and magnetic field, what
makes them experimentally accessible. Even though the quantum interference
itself does not effect the average value of DOS, the nontrivial corrections
to this quantity appear when, in addition, the interelectron interaction is
taken into account.

During mid 80's, the \emph{spatial variations} of the LDoS, conductivity and
other normal-metal properties have been revisited within the framework of the
mesoscopic physics.\cite{Mesosc} It was found that the quantum interference
effects also lead to appearance of nontrivial corrections to inhomogeneous
characteristics, for instance, to the LDoS correlation function.\cite{Lerner}
In contrast to the \textquotedblleft fast\textquotedblright\ atomic-scale
contribution of the Friedel oscillations, the latter manifest themselves in
smooth long-range spatial behavior of the LDoS correlation function as the
\textquotedblleft slow\textquotedblright\ power (or logarithmic in 2D case)
decay on the distances beyond the mean-free path $l$ and up to the
phase-breaking length $l_{\phi}\gg l$ (each of them is much larger than
interatomic distances). One can recognize the physical origin of such
phenomenon in spirit of the qualitative explanation of the weak localization
corrections given in terms of self-intersecting trajectories.\cite{WeakLocal}

The electron motion  in impure metal has the diffusive character. For every
pair of remote points $\mathbf{r}$ and $\mathbf{r} ^{\prime}$ with finite
probability one can find the self-intersecting quasiclassical trajectory which
starts from the point $\mathbf{r}$, passes close to the point
$\mathbf{r}^{\prime}$, and returns back to the initial point. An important
property is the existence of the opposite returning trajectory, which outcomes
from the point $\mathbf{r}^{\prime}$, passes close to the point $\mathbf{r}$,
and returns to the point $\mathbf{r}^{\prime}$ following roughly the same route
[see Fig.\ \ref{FigDiagram} (a)]. These two trajectories have two long joint
pieces where the electron motion is accompanied by the multiple
scattering on the same impurities. When time-inversion symmetry is present,
particles can move along these trajectories both in the same or in the opposite
directions. Looking at Fig.\ \ref{FigDiagram}(a) one can see that for each
trajectory there are the entry and the exit points of the joint routes (marked
by circles) separated by the distances $R_{1}$ and $R_{2}$ from the trajectory
origin. Existence of such diffusive trajectories leads to long-range
correlation of different properties, in particular, the local density of
states.

Quantitatively, this phenomenon can be described by the standard Green's
function diagrammatic technique. The diagram describing the long-range
correlations is shown in Fig. \ref{FigDiagram}(b). It contains either two
diffusons or two cooperons.\cite{Lerner} The two-diffuson diagram describes the
process in which the particles move in the same direction within the two joint
routes, while the two-cooperon diagram corresponds to the motion of particles
in the opposite directions. The Cooperon (diffuson) as the element of the
diagram describes the process of coherent scattering of electrons moving along
the joint routes. The blocks of three Green's functions (two retarded and one
advanced, or vice versa) are known as the Hikami boxes. They describe the
incoherent motion of electrons in the domains close to the entry and exit
points $\mathbf{r}$ and $\mathbf{r}^{\prime}$, where their routes divaricate.

Recently, STM measurements of the LDoS spatial variations have emerged as a new
powerful tool to characterize intrinsic inhomogeneities in impure
superconductors.\cite{STMcupr,STMpnict} These measurements revealed both rapid
oscillations with typical wave vectors connecting characteristic points at the
Fermi surface (the quasi-particle scattering interference patterns) and smooth
LDoS variations. In particular, studying the oscillating contribution provides
one of the ways to establish the symmetry of the superconducting order
parameter. The available theoretical description of these experiments is mostly
based on the single-impurity approximation \cite{CapriottiPRB03}, which becomes
insufficient at noticeable impurity concentration.

Our purpose in this article is to understand the behavior of inhomogeneous
LDoS of impure superconductors with different order parameter symmetries at
the length scales beyond the mean free path. We develop a theory which
properly accounts for the collective effects appearing during coherent
quasiparticle scattering on impurities. We demonstrate that the energy
dependence of the long-range correlation function of superconductor is indeed
sensitive to symmetry of the order parameter and nature of scatterers. The
inhomogeneous LDoS of superconductor can be directly mapped on that one of a
normal metal only in the simplest case of s-wave superconductor with elastic
scatterers. Presence of magnetic impurities or more nontrivial symmetry of the
order parameter results in the considerable complication of the LDoS
correlation function energy dependence while the spatial variations in all
cases do not change.

The article is organized in the following way. In Section II we start our
discussion introducing the Green's function formalism for study of the
inhomogeneous LDoS correlation function and refresh to a reader its properties
in normal metal. Section III is devoted to study of the LDoS correlation
function in s-wave superconductors and it consists of two subsections treating
cases with only elastic and both elastic and magnetic impurities. In Section
IV we consider the problem in the case of superconductor with d-wave symmetry
of the order parameter and elastic impurities. The cumbersome technical
details of calculation of the elements of diagrams for correlation function,
such as Hikami blocks, cooperons, and traces of large number of Pauli matrices
make the Appendices.

\section{Inhomogeneous LDoS in normal metals}

\begin{figure}[th]
\begin{center}
\includegraphics[width=2.5in]{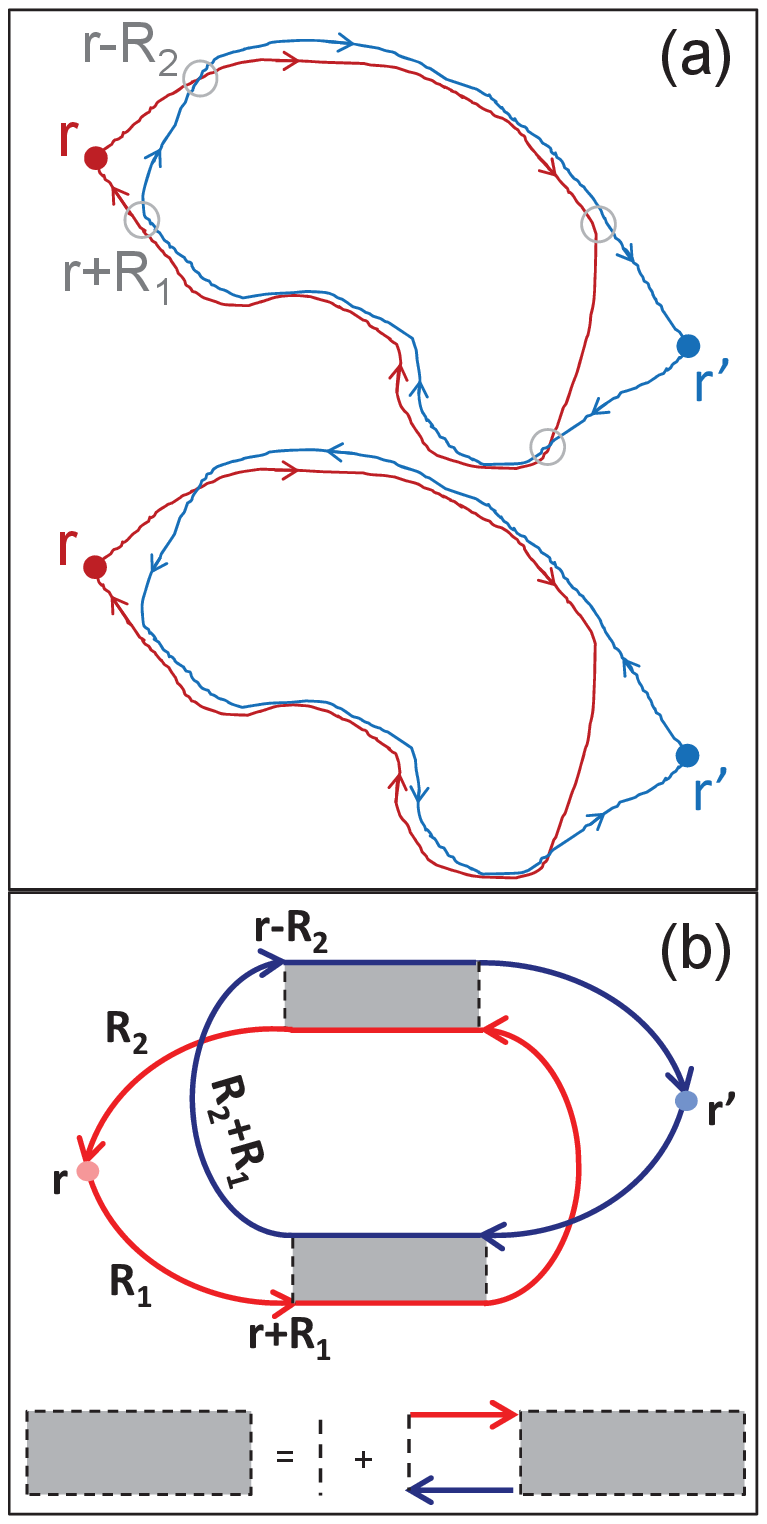}
\end{center}
\caption{(a) Returning quasiparticle trajectories leading to the long-range
correlations between the points $\mathbf{r}$ and $\mathbf{r}^{\prime}$. The
trajectories are characterized by two pieces of long joint routes within which
the diffusing particles move either in the same or in the opposite directions.
(b)Two-diffuson diagram describing long-range part of the LDoS correlation
function. Upper (lower) loop represents $G^{R}(E,\mathbf{r} ^{\prime
},\mathbf{r}^{\prime})$ ($G^{A}(E,\mathbf{r},\mathbf{r})$) correspondingly.
Shaded box represents diffuson. The two-cooperon contribution can be obtained
by reversing arrows direction in one of the loops.}
\label{FigDiagram}
\end{figure}
The case of normal metal represents a natural starting point and a convenient
reference. The effects of multiple scattering and quantum coherence on spatial
correlations of LDoS in disordered normal metals were considered in Ref.\
\onlinecite{Lerner}. It was found that two different contributions can be
distinguished in the LDoS correlation function. The short-range contribution
represents the Friedel oscillations modified by multiple impurity scattering.
The second long-range contribution appears due coherent diffusive propagation
of normal quasiparticles. In the following, we will focus namely on this
long-range term and, for completeness, we start with reproduction of its
calculation.

Our purpose is to evaluate the LDoS correlation function at the same energy
\begin{equation}
\mathcal{L}^{\left(  \mathrm{n}\right)}(\mathbf{r}-\mathbf{r}^{\prime},E)
=\left\langle N_{\mathrm{e}}^{\left(  \mathrm{n}\right)  }(E,\mathbf{r}
)N_{\mathrm{e}}^{\left(  \mathrm{n}\right)  }(E,\mathbf{r}^{\prime})\right\rangle
-\left\langle N_{\mathrm{e}}^{\left(  \mathrm{n}\right)  }(E)\right\rangle
^{2}.\label{LDoSCorrDef}
\end{equation}
As the LDoS per spin is related to the retarded Green's function
$G^{R}(E,\mathbf{r},\mathbf{r})$ by the standard relation,
\begin{equation}
N_{\mathrm{e}}^{\left(  \mathrm{n}\right)  }(E,\mathbf{r})=-\mathrm{Im}[G^{R}(E,\mathbf{r},\mathbf{r})]/\pi
,\label{NormDoS}
\end{equation}
this quantity can be expressed via retarded and advanced Green's functions:
\begin{align}
&\mathcal{L}^{\left(  \mathrm{n}\right)}(\mathbf{r}  -\mathbf{r}^{\prime},E)=\frac{1}{2\pi^{2}}\left\{
\mathrm{Re}\left[  \left\langle G^{R}(E,\mathbf{r},\mathbf{r})G^{A}
(E,\mathbf{r}^{\prime},\mathbf{r}^{\prime})\right\rangle \right.  \right.
\nonumber\\
&  \left.  \left.  -\left\langle G^{R}(E,0,0)\right\rangle \left\langle
G^{A}(E,0,0)\right\rangle \right]  \right. \label{LDoSCorrGreens}\\
&  \left.  -\!\mathrm{Re}\left[  \left\langle G^{R}(E,\mathbf{r}
,\mathbf{r})\ G^{R}(E,\mathbf{r}^{\prime},\mathbf{r}^{\prime})\right\rangle
\!-\!\left\langle G^{R}(E,0,0)\right\rangle ^{2}\right]  \right\}  ,\nonumber
\end{align}
where $\langle\ldots\rangle$ implies averaging over impurities distribution.
One can use this presentation for impurity averaging within the standard
Green's function approach. We will assume weak impurity scattering (Born
limit). The contributions to the LDoS correlation function can be represented
as diagrams consisting of two loops connected by impurity lines.

The long-range contribution to the LDoS correlation function is given by the
sum of the two-diffuson and two-cooperon diagrams, $\mathcal{L}^{\left(
\mathrm{n}\right)} =\mathcal{L}^{\left(
\mathrm{n}\right)}_{(2D)}+\mathcal{L}^{\left(  \mathrm{n}\right)}_{(2C)}$, see
Fig.\ \ref{FigDiagram}(b). Both of them give the same contributions, which can
be approximately written as
\begin{equation}
\mathcal{L}^{\left(  \mathrm{n}\right)}_{(2D)}(\mathbf{r}\!-\!\mathbf{r}^{\prime},E)
=\mathcal{L}^{\left(  \mathrm{n}\right)}_{(2C)}(\mathbf{r}\!-\!\mathbf{r}^{\prime},E)\approx\frac{\left\vert
\mathcal{B}\left(  E\right)  \right\vert ^{2}}{2\pi^{2}}C_{\mathrm{n}}^{2}(\mathbf{r}
\!-\!\mathbf{r}^{\prime}),\label{TwoCoopNormal}
\end{equation}
where $C_{\mathrm{n}}(\mathbf{r}-\mathbf{r}^{\prime})$ is the cooperon,
\begin{equation}
\mathcal{B}^{\mathrm{(n)}}\left(  E\right)  =\int d\mathbf{R}_{1}\int d\mathbf{R}_{2}\bar
{G}^{R}(\mathbf{R}_{1})\bar{G}^{A}(\mathbf{R}_{1}\mathbf{-\mathbf{R}}_{2}
)\bar{G}^{R}(\mathbf{R}_{2})\label{HikamiNormal}
\end{equation}
is the ``Hikami box'', and
\[
\bar{G}^{R,A}(\mathbf{R})\!\equiv\!\left\langle G^{R,A}(E,\mathbf{r}
,\mathbf{r}\!+\!\mathbf{R})\right\rangle \!=\!\int\frac{d^{D}\mathbf{p}}{(2\pi)^{D}
}\frac{\exp(i\mathbf{pR})}{E\!-\!\epsilon_{\mathbf{p}}\pm i/2\tau}
\]
are the averaged Green's functions with $\mathrm{D}$ being the space
dimensionality. Here $\tau$ is the elastic scattering time,
$\epsilon_{\mathbf{p}}=v_{F}(p-p_{F})$, $v_{F}$ and $p_{F}$ are the Fermi
velocity and momentum. Integration in Eq.\ (\ref{HikamiNormal}) gives a very
simple result
\begin{equation}
\mathcal{B}^{\mathrm{(n)}}\left(  E\right)  =2\pi i\nu\tau^{2}
\label{HikamiNormalResult}
\end{equation}
with $\nu=\langle N_{\mathrm{e}}^{\left(  \mathrm{n}\right)  }(E,\mathbf{r})\rangle$ as the average value of LDoS.

Hence, the long-range behavior of the LDoS is completely determined by the
cooperon $C_{\mathrm{n}}(\mathbf{r}-\mathbf{r}^{\prime})$ whose Fourier transform is well
known \cite{WeakLocal}
\begin{equation}
C_{\mathrm{n}}(q)=\frac{1}{2\pi\nu\tau}f_{\mathrm{D}}(lq)\ \text{with }f_{\mathrm{D}
}(lq)\approx\frac{\mathrm{D}}{l^{2}q^{2}}\text{ for }lq\ll1,\label{NormCoop-q}
\end{equation}
where $l=v_{F}\tau$ is the mean-free path. In real space
\begin{equation}
C_{\mathrm{n}}(r)=\frac{\mathrm{D}}{2\pi\nu\tau l^{\mathrm{D}}}\tilde{f}_{\mathrm{D}
}(r/l).\label{CooperonNormal-r}
\end{equation}
At $r\gg l$
\begin{equation}
\tilde{f}_{\mathrm{D}}(r)\approx\frac{1}{\pi}\left\{
\begin{array}
[c]{cc}
\ln(l_{\phi}/r) & \text{ for }\mathrm{D}=2\\
3/4r & \text{for }\mathrm{D}=3
\end{array}
\right.  .\label{fC-r}
\end{equation}
For the 2D case the logarithmic divergency is cut off at $q\sim1/l_{\phi}$,
where $l_{\phi}$ is the phase-decoherence length. As the Green's functions
decay at the distances of the order of the mean-free path, in order to
evaluate the long-range behavior $|\mathbf{r}-\mathbf{r}^{\prime}|\gg l$, we
replaced in Eq.\ (\ref{TwoCoopNormal}) arguments of both cooperons with
$|\mathbf{r}-\mathbf{r}^{\prime}|$.

Substituting the results (\ref{HikamiNormalResult}) and
(\ref{CooperonNormal-r}) into Eq.\ (\ref{TwoCoopNormal}), we obtain the
long-range asymptotic expression of the LDoS correlation function
\cite{Lerner}
\begin{align}
\mathcal{L}^{\left(  \mathrm{n}\right)}(\mathbf{r}-\mathbf{r}^{\prime},E) &  =\left(
2\nu\tau^{2}\right)  ^{2}C^{2}_{\mathrm{n}}(\mathbf{r}-\mathbf{r}^{\prime})\nonumber\\
&  =\nu^{2}\frac{a_{\mathrm{D}}}{\left(  k_{F}l\right)  ^{2\mathrm{D}-2}
}f_{\mathrm{D}}^{2}(|\mathbf{r}-\mathbf{r}^{\prime}|/l)\label{DoSCorrNormal}
\end{align}
with $a_{2}=4$ and $a_{3}=4\pi^{2}$. From this result we see that LDoS
variations are small in comparison with the average DoS by the parameter
$\left( k_{F}l\right)  ^{\mathrm{D}-1}$. It is important to stress that in
contrast to the short-ranged Friedel oscillations, they decay slowly at large
distances up to $l_{\phi}$, as $\ln^{2}(l_{\phi}/r)$ for the 2D case and as
$1/r^{2}$ for the 3D case.

\section{s-wave superconductors}

\subsection{Potential impurities}

We start consideration of superconducting state with the simplest situation of
purely potential scattering and s-wave symmetry of the order parameter. In this
case the inhomogeneous LDoS is directly related to its normal-state
counterpart. Indeed, using eigenstates expansion, the normal-state LDoS can be
represented as
\begin{equation}
N_{\mathrm{e}}^{\left(  \mathrm{n}\right)  }(E,\mathbf{r})=\sum_{i}|\psi
_{i}(\mathbf{r})|^{2}\delta(E-E_{i}),\label{NDoSEigen}
\end{equation}
where $\psi_{i}(\mathbf{r})$ are eigenfunctions and $E_{i}$ are eigenenergies
of the quasiparticle states. In case of potential scattering, the corresponding
eigenenergies in superconducting state become $\pm\sqrt {E_{i}^{2}+\Delta^{2}}$
where $\Delta$ is the gap parameter, and the two-component Bogoliubov wave
function of quasiparticle state in superconductor
$(U_{i,\pm}(\mathbf{r}),V_{i,\pm}(\mathbf{r}))$ is proportional to the normal
state wave function $(U_{i,\pm}(\mathbf{r}),V_{i,\pm}
(\mathbf{r}))=(u_{i,\pm},v_{i\,,\pm})\psi_{i}(\mathbf{r}),$ where $u_{i,\pm}$
and $v_{i,\pm}$ are coordinate-independent constants $u_{i,\alpha}=\left(
\alpha/\sqrt{2}\right)  \sqrt{1+\alpha E_{i}/\sqrt{E_{i}^{2}+\Delta^{2}}}$,
$v_{i,\alpha}=\left(  1/\sqrt{2}\right)  \sqrt{1-\alpha E_{i}/\sqrt{E_{i}
^{2}+\Delta^{2}}}$, $|u_{i,\pm}|^{2}+|v_{i,\pm}|^{2}=1$.\cite{deGennes}  A
quantity commonly evaluated for superconductors is the density of states
\emph{for excitations} which in normal state corresponds to symmetric
combination $N_{\mathrm{ex}}^{\left(  \mathrm{n}\right)  }(E,\mathbf{r}
)=N_{\mathrm{e}}^{\left(  \mathrm{n}\right)  }(E,\mathbf{r})+N_{\mathrm{e}
}^{\left(  \mathrm{n}\right)  }(-E,\mathbf{r})$. The excitation LDoS in
superconducting state, $N_{\mathrm{ex}}^{\left(  \mathrm{s}\right)
}(E,\mathbf{r})$, can be represented in the form of eigenstate expansion as
\begin{align}
&N_{\mathrm{ex}}^{(\mathrm{s})}(E,\mathbf{r}) \!=\!\!\!\!\sum
_{i,\alpha=\pm1}\!\!\!\left[  |U_{i,\alpha}(\mathbf{r})|^{2}\!+\!|V_{i,\alpha
}(\mathbf{r})|^{2}\right]\! \delta\!\left(\!E\!-\!\alpha\sqrt{E_{i}^{2}\!+\!\Delta^{2}
}\right)\nonumber\\
&  =\sum_{i,\alpha=\pm1}|\psi_{i}(\mathbf{r})|^{2}\frac{|E|}{\sqrt
{E^{2}\!-\!\Delta^{2}}}\delta\left(E_{i}\!-\!\alpha\sqrt{E^{2}-\Delta^{2}}\right)\nonumber\\
&  =\frac{|E|}{\sqrt{E^{2}-\Delta^{2}}}N_{\mathrm{ex}}^{\left(  \mathrm{n}
\right)  }(\sqrt{E^{2}-\Delta^{2}},\mathbf{r})\label{SDoSEigen}
\end{align}
(in the second line we change the variable of the $\delta$-function). Such a
simple connection between the normal and superconducting LDoS is one of
consequences of the Anderson theorem and it provides the following relation
between the normal-state and superconducting LDoS correlation functions
\begin{equation}
\mathcal{L}^{\mathrm{(s)}}_{\mathrm{ex}}(E,\mathbf{r})/{[\nu_{\mathrm{ex}}^{\mathrm{(s)}}(E)]^{2}}=\mathcal{L}
^{\mathrm{(n)}}_{\mathrm{ex}}(\sqrt{E^{2}-\Delta^{2}},\mathbf{r})/(2\nu
)^{2},\label{SDoSCorr}
\end{equation}
where $\nu_{\mathrm{ex}}^{\mathrm{(s)}}(E)=2\nu E/\sqrt{E^{2}-\Delta^{2}}$ is
the average superconducting DoS for excitations and
$\mathcal{L}^{\mathrm{(n)}}_{\mathrm{ex}}(E,\mathbf{r})\approx\mathcal{L}^{\mathrm{(n)}}(E,\mathbf{r}
)+\mathcal{L}^{\mathrm{(n)}}(-E,\mathbf{r})$. Even though in the following we
only consider $\mathcal{L}^{\mathrm{(s)}}_{\mathrm{ex}}(E,\mathbf{r})$, for
completeness we also present a useful general relation for the
\emph{electronic} LDoS,
\begin{align}
&N_{\mathrm{e}}^{\left(  \mathrm{s}\right)  }(E,\mathbf{r})  =\sum
_{i,\alpha=\pm1}\!|U_{i\,,\alpha}(\mathbf{r})|^{2}\delta\left(E-\alpha\sqrt{E_{i}
^{2}+\Delta^{2}}\right)\nonumber\\
&  =\sum_{\alpha=\pm1}\frac{E\!+\!\alpha\sqrt{E^{2}\!-\!\Delta^{2}}}{2\sqrt
{E^{2}-\Delta^{2}}}N_{\mathrm{e}}^{\left(  \mathrm{n}\right)  }(\alpha
\sqrt{E^{2}\!-\!\Delta^{2}},\mathbf{r})
\end{align}
In particular, this general relation allows immediately to reproduce the
single-impurity result reported for the s-wave case in Ref.\
\onlinecite{CapriottiPRB03}.

Even though the long-range tail of LDoS for disordered s-wave superconductors
can be immediately obtained from the normal-state result, it is useful,
nevertheless, to rederive it formally, within the Green's function approach.
This will allow us to generalize such an approach later for less trivial
situations for which the above argument does not work any more.

The long-range contribution to the LDoS correlation function is again
determined by the diagram shown in Fig. \ref{FigDiagram}(b), but both the
Green's functions and cooperons now have the matrix structure. In order to
calculate them, it is convenient to use Nambu formalism decomposing the Green's
functions over Pauli matrices $\hat{\tau}^{j}$ and the cooperons over the
direct product $\hat{\tau}^{k}\otimes\hat{\tau}^{k^{\prime}}$ of them
\begin{subequations}
\begin{align}
G^{R,A}  &  \rightarrow G_{\alpha\beta}^{R,A}=g_{j}^{R,A}\tau_{\alpha\beta
}^{j}\label{GreenPauli}\\
C^{\left(  \mathrm{s}\right)  }  &  \rightarrow C^{\left(  \mathrm{s}\right)}_{\alpha\beta,\gamma\delta}
=C^{\left(  \mathrm{s}\right)}_{kk^{\prime}}\tau_{\alpha\beta}^{k}\tau_{\gamma\delta}^{k^{\prime}}
\label{CooperonPauli}
\end{align}
\end{subequations}
We assume summation with respect to repeated indices. Let us note that the
Cooperon $4\times4$ matrix $C^{\left(  \mathrm{s}\right)}_{kk^{\prime}}$ in
fact has the $2\times2$ block structure\cite{Kulik,VarlamovDorinZhETF86}:
\begin{equation}
C^{\left(  \mathrm{s}\right)}_{kk^{\prime}}=\left[
\begin{array}
[c]{cc}
\widehat{C}^{\left(  \mathrm{s}\right)}_{A} & 0\\
0 & \widehat{C}^{\left(  \mathrm{s}\right)}_{B}
\end{array}
\right]  .\label{cooperonblock}
\end{equation}

For potential scattering the averaged superconducting Green's functions are
given by
\begin{align}
\hat{G}^{R,A}\left(  E,\mathbf{p}\right)   &  =\frac{\alpha^{R,A}\left(
E\hat{\tau}^{0}+\Delta\hat{\tau}^{1}\right)  +\epsilon_{\mathbf{p}}\hat{\tau
}^{3}}{(\alpha^{R,A})^{2}\left(  E^{2}-\Delta^{2}\right)  -\epsilon
_{\mathbf{p}}^{2}},\label{AvSupGreen}\\
\alpha^{R,A}  &  =1\mp\frac{i}{2\tau\sqrt{E^{2}-\Delta^{2}}}.\nonumber
\end{align}
and for real $\Delta$ they do not contain $\hat{\tau}^{2}$ component.

In this formalism it is convenient to deal with LDoS for excitations
$N_{\mathrm{ex}}^{\left(  \mathrm{s}\right)  }$ which is related to the trace
of the Green's function
\begin{equation}
N_{\mathrm{ex}}^{\left(  \mathrm{s}\right)  }(E,\mathbf{r})=-\frac{1}{\pi
}\mathrm{Im}\left[  \mathrm{Tr}\ \hat{G}^{R}\left(  E,\mathbf{r}
,\mathbf{r}\right)  \right]  .\label{SupLDoSDef}
\end{equation}
Comparing this equation with the previous normal state LDoS definition
(\ref{NormDoS}) one can see that $N_{\mathrm{ex}}^{\left(  \mathrm{s}\right)
}(E,\mathbf{r})$ has an additional factor two since it contains both electron
and hole contributions. In particular, the average LDoS for excitations is
given by $\nu_{\mathrm{ex}}^{\mathrm{(s)}}$.
The corresponding expression for the two-cooperon diagram can be represented using the Pauli-matrices decomposition:
\begin{subequations}
\begin{align}
& \mathcal{L}^{\mathrm{(s)}}_{\mathrm{ex}(2C)}=\frac{4}{2\pi^{2}}U_{km}U_{k^{\prime}m^{\prime}}^{\ast
}C^{\left(  \mathrm{s}\right)}_{kk^{\prime}}C^{\left(  \mathrm{s}\right)}_{mm^{\prime}},\label{TwoCoopSC}\\
&  U_{km}=\frac{1}{2}\mathrm{Tr}\left(  \hat{\tau}^{i}\hat{\tau}^{k}\hat{\tau
}^{n}\hat{\tau}^{m}\hat{\tau}^{j}\right)  \mathcal{B}_{inj}^{\left(
\mathrm{s}\right)  },\label{VertexSC}\\
& \mathcal{B}_{inj}^{\left(  \mathrm{s}\right)  }=\!\int \!d\mathbf{R}_{1}\!\int
\!d\mathbf{R}_{2}g_{i}^{A}(\mathbf{R}_{1})g_{n}^{R}(\mathbf{R}_{1}\!
-\!\mathbf{R}_{2})g_{j}^{A}(\mathbf{R}_{2}).\label{Hikami-swave}
\end{align}
\end{subequations}

The computational details of superconducting cooperon components are presented
in Appendix \ref{App-Cooperon-s}. It turns out that its singular components
are related to the normal-state cooperon (\ref{NormCoop-q}) as
\begin{subequations}
\begin{align}
C^{\left(  \mathrm{s}\right)}_{ij}(\mathbf{q})  &  =\frac{E^{2}}{E^{2}\!-\!\Delta^{2}}\left(
-\frac{\Delta}{E}\right)  ^{i+j}\frac{C_{\mathrm{n}}(\mathbf{q})}{2}\text{, for
}i,j=0,1\text{ }\label{SupCoop01}\\
C^{\left(  \mathrm{s}\right)}_{33}(\mathbf{q})  &  =C_{\mathrm{n}}(\mathbf{q}) /2\label{SupCoop3}
\end{align}
\label{SupCoop-s}
\end{subequations}
One can see that in the limit of normal metal, $\Delta\rightarrow 0$, the only
nonzero components remained are $C^{\left(  \mathrm{s}\right)}_{00}$ and
$C^{\left( \mathrm{s}\right)}_{33}$.

Computing the trace of five Pauli matrices in Eq.\ (\ref{VertexSC}), taking into
account (i) symmetry of $\mathcal{B}_{inj}^{\left(  \mathrm{s}\right)  }$ with
respect to indices $i$ and $j$ and (ii) the absence of $\hat{\tau}^{2}$ components
in all decompositions (see Appendix \ref{App-trace-Pauli}), we obtain
\begin{align*}
U_{km} &  =\delta_{m0}\mathcal{B}_{iki}^{\left(  \mathrm{s}\right)  }
+\delta_{k0}\mathcal{B}_{imi}^{\left(  \mathrm{s}\right)  }+\left(
\delta_{mk}-2\delta_{m0}\delta_{k0}\right)  \mathcal{B}_{i0i}^{\left(
\mathrm{s}\right)  }\\
&  +2\left\{  (1-\delta_{k0})\mathcal{B}_{0mk}^{\left(  \mathrm{s}\right)
}+(1-\delta_{m0})\mathcal{B}_{0km}^{\left(  \mathrm{s}\right)  }\right. \\
&  \left.  -(1-\delta_{n0})\mathcal{B}_{0nn}^{\left(  \mathrm{s}\right)
}\left(  \delta_{km}-2\delta_{k0}\delta_{m0}\right)  \right\}  ,
\end{align*}
where summation with respect to the index $i=0,1,3$ is assumed in the first
three terms. The components of the Hikami boxes are computed in Appendix
\ref{App-Hikami-s}. For $i,j,m=0,1$
\[
\mathcal{B}_{imj}^{\left(  \mathrm{s}\right)  }\!=\!-\frac{i\pi\nu\tau^{2}}
{2}\frac{E^{3}\left(  \sqrt{E^{2}\!-\!\Delta^{2}}+\frac{3i}{2\tau}\right)
}{\left(  E^{2}\!-\!\Delta^{2}\right)  ^{3/2}\!\left(  \sqrt{E^{2}
\!-\!\Delta^{2}}+\frac{i}{2\tau}\right)  }\left(  \frac{\Delta}{E}\right)
^{i+j+m}\text{ }
\]
and
\begin{align*}
\mathcal{B}_{033}^{\left(  \mathrm{s}\right)  } &  =-\frac{i\pi\nu E\tau^{2}
}{2\sqrt{E^{2}-\Delta^{2}}},\\
\mathcal{B}_{303}^{\left(  \mathrm{s}\right)  } &  =-\frac{i\pi\nu E\tau^{2}
}{2\sqrt{E^{2}-\Delta^{2}}}\frac{\sqrt{E^{2}-\Delta^{2}}-\frac{i}{2\tau}
}{\sqrt{E^{2}-\Delta^{2}}\!+\!\frac{i}{2\tau}}.
\end{align*}
Due to the specifics of the cooperon $C^{\left(  \mathrm{s}\right)}_{ij}(\mathbf{q})$ structure for
$i,j=0,1$, seen from Eq.\ (\ref{SupCoop01}), we need only the combination
$U_{00}+U_{11}\left(  \Delta/E\right)  ^{2}-2\left(  \Delta/E\right)  U_{01}$
for which we find a very simple relation
\[
U_{00}+U_{11}\left(  \Delta/E\right)  ^{2}-2\left(  \Delta/E\right)
U_{01}=U_{33}\left[  1-\left(  \Delta/E\right)  ^{2}\right]
\]
and for $U_{33}$ we derive
\[
U_{33}=-2i\pi\nu\tau^{2}\frac{E}{\sqrt{E^{2}-\Delta^{2}}}.
\]
Collecting terms, we finally obtain for the \emph{total} correlation function,
$\mathcal{L}^{\mathrm{(s)}}_{\mathrm{ex}}=\mathcal{L}^{\mathrm{(s)}}_{\mathrm{ex}(2C)}+\mathcal{L}
^{\mathrm{(s)}}_{\mathrm{ex}(2D)}$,
\begin{align}
&  \mathcal{L}^{\mathrm{(s)}}_{\mathrm{ex}}(\mathbf{r}-\mathbf{r}^{\prime},E)=\frac{4}
{\pi^{2}}|U_{33}|^{2}\nonumber\\
&  \times\left[  \left(  1-\left(  \frac{\Delta}{E}\right)  ^{2}\right)
^{2}[C^{\left(  \mathrm{s}\right)}_{00}(\mathbf{r}-\mathbf{r}^{\prime})]^2
+[C^{\left( \mathrm{s}\right)}_{33}(\mathbf{r}
-\mathbf{r}^{\prime})]^2\right] \nonumber\\
&  =2\left(  \frac{2\nu\tau^{2}E}{\sqrt{E^{2}-\Delta^{2}}}\right)  ^{2}
C^{2}_{\mathrm{n}}(\mathbf{r}-\mathbf{r}^{\prime}).\label{LSup-el}
\end{align}
This result explicitly confirms the relation (\ref{SDoSCorr}) based on
Anderson-theorem arguments.

\subsection{Magnetic impurities}

In this section we evaluate the LDoS correlation function for s-wave
superconductor with magnetic scatterers. Since the seminal paper of Abrikosov
and Gor'kov \cite{AG60} it is established that the magnetic impurities
dramatically suppress superconductivity and strongly influence the
quasiparticle density of states. The original mean-field approach (AG theory)
suggested that the hard gap in the average DoS is not eliminated by magnetic
impurities. This hard gap decreases with increasing the magnetic-impurities
concentration and vanishes at certain critical concentration so that the
gapless superconducting state exists within small range of concentrations. More
accurate later treatments beyond the mean-field approach
\cite{Simons01,ShytovPRL03} have demonstrated that the low-energy quasiparticle
states are in fact generated for all magnetic-impurities concentrations meaning
that, strictly speaking, the hard gap is eliminated by any amount of magnetic
impurities. For small concentrations, however, the average DoS at low energies
have exponentially small tail.

Formally, in the presence of magnetic scattering the Green's function can be
still presented in the form similar to Eq.\ (\ref{AvSupGreen}), but within the
mean-field approach the renormalized energy and gap now are determined from the
self-consistent transcendental equations.\cite{AG60} This result is usually
obtained taking into account the spin structure of the superconductive Green's
function \cite{Maki}, which means the additional increase of the matrix
dimensionality to $4\times4$. In the calculation of the two-cooperon diagrams in
Nambu representation (\ref{TwoCoopSC}) the number of Pauli matrices in traces
is already large. That is why, for the sake of simplicity, we will consider
below magnetic scatterers as Ising impurities oriented in the same direction.
This allows us to preserve $2\times2$ matrix structure of the Green's function,
which significantly simplifies calculations. The physical picture remains
practically identical to the case of isotropic magnetic impurities. Even for
such a minimum model describing the pair-breaking scattering, the calculations
and results become rather cumbersome. Technically similar work has been done in
Ref.\ \onlinecite{SmithAmbegPRB00} where suppression of the transition
temperature by magnetic impurities in combination with Coulomb effects has been
investigated. This thermodynamic problem required summation of ladder diagrams
with elastic and magnetic impurity lines in the Matsubara-frequency
presentation. In principle, the cooperon components at real energies could be
obtained from the results of this work via analytic continuation. However, this
procedure is not trivial at all.

The long-range LDoS correlation function is still defined by
Eq.\ (\ref{TwoCoopSC}) but the Green's functions, cooperons, and Hikami boxes
are considerably modified by magnetic scattering. Detailed derivations of
these objects are presented in Appendices \ref{App-Cooperon-sm} and
\ref{App-Hikami-sm}. Instead of Eq.\ (\ref{AvSupGreen}), Green's functions are
given by
\begin{equation}
\hat{G}^{R,A}\left(  \mathbf{p}\right)  =\frac{\tilde{E}_{\pm}\hat{\tau}
^{0}+\tilde{\Delta}_{\pm}\hat{\tau}^{1}+\epsilon_{\mathbf{p}}\hat{\tau}^{3}
}{ \tilde{E}_{\pm}^{2}-\tilde{\Delta}_{\pm}
^{2}-\epsilon_{\mathbf{p}}^{2}}\label{GRAmp}
\end{equation}
with the renormalized energy and gap
\begin{subequations}
\begin{align}
\tilde{E}_{\pm} &  =E\pm\frac{i\eta_{\pm}}{2\tau_{0}\sqrt{\eta_{\pm}^{2}-1}
},\label{ERAmp}\\
\tilde{\Delta}_{\pm} &  =\Delta\pm\frac{i}{2\tau_{1}\sqrt{\eta_{\pm}^{2}-1}
},\label{DRAmp}
\end{align}
where $1/\tau_{\alpha}=1/\tau+(-1)^{\alpha}/\tau_{m}$, $\tau$ and $\tau_{m}$
are the potential and magnetic scattering times. Here and below the subscript
``$+$'' (``$-$'') corresponds to the retarded (advanced) components.
The parameter $\eta_{\pm
}=\tilde{E}_{\pm}/\tilde{\Delta}_{\pm}$ has to be determined from the
equation
\end{subequations}
\begin{equation}
\eta_{\pm}\left(  1\mp\frac{i}{\tau_{m}\Delta\sqrt{\eta_{\pm}^{2}-1}}\right)
=\frac{E}{\Delta}.\label{SelfConsuRA}
\end{equation}
The average DoS is connected with $\eta_{+}$ by relation \cite{AG60}
\begin{equation}
\nu_{\mathrm{ex}}^{(\mathrm{sm})}(E)=2\nu\operatorname{Re}\frac{\eta_{+}}
{\sqrt{\eta_{+}^{2}-1}}.
\label{AvDoSMP}
\end{equation}

The $2\times2$ blocks of the cooperon Eq.\ (\ref{cooperonblock}) computed in
Appendix \ref{App-Cooperon-sm} are
\begin{align}
\hat{C}_{A}^{\mathrm{(sm)}}=&\frac{1/4\pi\nu}{1-\tilde{\gamma}_{\mathbf{q}}}\left[  \left(  \frac{1}{\tau_{m}}
\!+\!\frac{\chi_1\tilde{\gamma}_{\mathbf{q}}}{2\tau^{\ast}}
\right)  \hat{\tau}^{0} \right.\nonumber\\
&\left.+\left(  \frac{1}{\tau}
\!-\!\frac{\tilde{\gamma}_{\mathbf{q}}}{2\tau^{\ast}}\right)  \hat{\tau}^{3}\!-\!\frac
{\tilde{\gamma}_{\mathbf{q}}}{\tau^{\ast}}\frac{\operatorname{Re}\eta_{+}}{\left\vert \eta_{+}
^{2}-1\right\vert }\hat{\tau}^{1}\right]
\end{align}
and
\begin{align}
\hat{C}_{B}^{\mathrm{(sm)}}=&\frac{1/4\pi\nu}{1-\tau_{\mathrm{qp}}\tilde{\gamma}_{\mathbf{q}}\left(  \frac{1}{\tau}
+\frac{\chi_2}{\tau_{m}}\right)  }\left[  \left(  \frac{1}{\tau_{m}}
+\frac{\chi_2\tilde{\gamma}_{\mathbf{q}}}{2\tau^{\ast}}\right)  \hat{\tau}^{0}\right.\nonumber\\-&\left.\left(
\frac{1}{\tau}-\frac{\tilde{\gamma}_{\mathbf{q}}}{2\tau^{\ast}}\right)  \hat{\tau}
^{3}+i\frac{\tilde{\gamma}_{\mathbf{q}}}{\tau^{\ast}}\frac{\operatorname{Im}\eta_{+}
}{\left\vert \eta_{+}^{2}-1\right\vert }\hat{\tau}^{2}\right]
\end{align}
with
\begin{align*}
\tau_{\mathrm{qp}}^{-1} &  =\frac{1}{\tau}+\frac{\chi_1}{\tau_{m}},\ \chi_1=\frac{|\eta_{+}
|^{2}+1}{\left\vert \eta_{+}^{2}-1\right\vert },\ \chi_2=\frac{|\eta_{+}
|^{2}-1}{\left\vert \eta_{+}^{2}-1\right\vert },\ \\
\tilde{\gamma}_{\mathbf{q}} &  =\left\langle \frac{1}{1+\tau_{\mathrm{qp}}^{2}\left(
\mathbf{v}_{F}\mathbf{q}\right)  ^{2}}\right\rangle ,\;\frac{1}{\tau^{\ast}
}=\tau_{\mathrm{qp}}\left(  \frac{1}{\tau^{2}}-\frac{1}{\tau_{m}^{2}}\right)  .
\end{align*}
Note that the component $\hat{C}_{A}^{\mathrm{(sm)}}$ still has diffusive
divergency for $q\rightarrow 0$, while in $\hat{C}_{B}^{\mathrm{(sm)}}$ it is cut
off by magnetic scattering. Finally, we compute the Hikami-box components
\begin{align*}
\mathcal{B}_{000}^{\left(  \mathrm{sm}\right)  } &  =-\frac{i\pi\nu\tau
_{\mathrm{qp}}^{2}}{2}\frac{\tilde{E}_{+}^{2}\tilde{E}_{-}\left(  \sqrt{\tilde{E}
_{+}^{2}-\tilde{\Delta}_{+}^{2}}+\frac{i}{\tau_{\mathrm{qp}}}\right)  }{\left[
\tilde{E}_{+}^{2}-\tilde{\Delta}_{+}^{2}\right]  ^{3/2}
\sqrt{\tilde{E}_{-}^{2}-\tilde{\Delta}_{-}^{2}}},\\
\mathcal{B}_{klm}^{\left(  \mathrm{sm}\right)  } &  =\mathcal{B}
_{000}^{\left(  \mathrm{sm}\right)  }\zeta_{+}^{k+m}\zeta_{-}
^{l},\ \text{for\ }k,l,m=0,1
\end{align*}
with notations $\zeta_{+}=\tilde{\Delta}_{+}/\tilde{E}_{+}=1/\eta_{+}$ and
\begin{align*}
\mathcal{B}_{k33}^{\left(  \mathrm{sm}\right)  } &  =-\frac{i\pi\nu\tau
_{\mathrm{qp}}^{2}}{2}\frac{\tilde{E}_{+}\zeta_{+}^{k}}{\sqrt{\tilde{E}_{+}
^{2}-\tilde{\Delta}_{+}^{2}}}=-\frac{i\pi\nu\tau_{\mathrm{qp}}^{2}}{2}\frac{\eta
_{+}\zeta_{+}^{k}}{\sqrt{\eta_{+}^{2}-1}}\\
\mathcal{B}_{3k3}^{\left(  \mathrm{sm}\right)  } &  =-\frac{i\pi\nu\tau
_{\mathrm{qp}}^{2}}{2}\frac{\tilde{E}_{-}\zeta_{-}^{k}}{\sqrt{\tilde{E}_{+}
^{2}-\tilde{\Delta}_{+}^{2}}}
\end{align*}
for $k=0,1$.

Taking into account all these modifications, one can calculate the LDoS
correlation function in the presence of magnetic impurities. Its long-ranged part
comes from the A-block only (as it was already mentioned, the divergence at
$q\rightarrow0$ in $\widehat{C}_{B}$ is cut off by paramagnetic impurities,
see Eq.(\ref{CB}) ) and the final answer can be represented as
\begin{equation}
\mathcal{L}^{\mathrm{(sm)}}_{\mathrm{ex}}(\mathbf{r}
-\mathbf{r}^{\prime},E)=\left(  2\nu\tau_{0}^{2}\right)  ^{2}\left\vert
\mathcal{R}\right\vert ^{2}C^{2}_{\mathrm{n}}(\mathbf{r}-\mathbf{r}^{\prime}
),\label{LDoSCorrMPLong}
\end{equation}
where $1/\tau_{0}=1/\tau+1/\tau_{m}$,
$C_{\mathrm{n}}(\mathbf{r}-\mathbf{r}^{\prime})$ is the normal-state cooperon
(\ref{NormCoop-q}) and
\begin{widetext}
\begin{align}
&  \mathcal{R}=\frac{\chi_1+1}{4}\frac{1+\tau/\tau_{m}}{1+\chi_1\tau/\tau_{m}} \label{RLDoSMPLong}\\
&  \times\left\{  \frac{1+\vartheta^{2}-2\vartheta\zeta_{-}}{\sqrt
{1-\zeta_{-}^{2}}}+\frac{1-\vartheta^{2}}{\sqrt{1-\zeta_{+}^{2}}}
+\frac{2\zeta_{+}}{1-\zeta_{+}^{2}}\left(  \frac{2}{\sqrt{1-\zeta_{-}^{2}
}}-\frac{\tilde{E}_{-}}{\tilde{E}_{+}}\frac{1}{\sqrt{1-\zeta_{+}^{2}}}\right)  \left[
\left(  1+\vartheta^{2}\right)  \operatorname{Re}[\zeta_{+}]-\vartheta\left(
1+\zeta_{+}\zeta_{-}\right)  \right]  \right\}\nonumber
\end{align}
\end{widetext}
with
\[
\vartheta=r\sqrt{\frac{\chi_1-1}{\chi_1+1}},\ \frac{\tilde{E}_{-}}{\tilde{E}_{+}}
=\frac{-\frac{E}{\Delta}r+\eta_{-}}{-\frac{E}{\Delta}r+\eta_{+}}
,\ r=\frac{1-\tau/\tau_{m}}{1+\tau/\tau_{m}}.
\]
\begin{figure}[ptb]
\begin{center}
\includegraphics[width=2.5in]{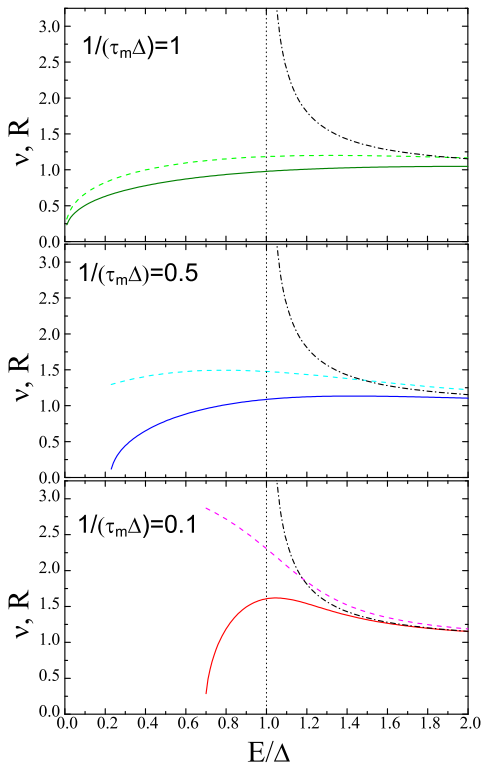}
\end{center}
\caption{
Examples of the energy dependences of the normalized average DoS $\nu_{\mathrm{ex}}^{(\mathrm{sm})}(E)/(2\nu)$,
Eq.\ (\ref{AvDoSMP}), (solid lines),
and function $|\mathcal{R}(E)|$, Eq.\ (\ref{RLDoSMPLong}), (dashed lines), which
determines the long-range LDoS correlations for $1/(\tau\Delta)=5$ and
different values of $1/(\tau_{m}\Delta)$. Dash-dotted lines show for reference
the BCS density of states.} \label{RLDoSMag}
\end{figure}
Normalization of the dimensionless function $\mathcal{R}(E)$ is selected by the
condition $\mathcal{R}(E)\rightarrow1$ for $E\rightarrow\infty$. This function
together with normalized average DoS is plotted in Fig.\ \ref{RLDoSMag} for
several values of the pair-breaking parameter $1/(\tau_{m}\Delta)$. As one can
see, the energy dependence of the correlation function described by
$\mathcal{R}(E)$ is quite different from that one of average density of states.
The most dramatic disparity is observed at small concentration of magnetic
impurities, for $1/(\tau_{m}\Delta)=0.1$. In this case the function
$\mathcal{R}(E)$ monotonically increases with decrease of energy, while
$\nu(E)$ goes down and finally vanishes at $E\approx 0.7 \Delta$. Here, at the
edge of  local density of states gap, the function $\mathcal{R}(E)$ reaches its
maximum. With increase of magnetic impurities concentration, when
$1/(\tau_{m}\Delta)=0.5$, the function $\mathcal{R}(E)$ still passes noticeably
above $\nu(E)$, reaches its weakly pronounced maximum and remains finite when
$\nu(E)$ turns zero. The behavior of both functions becomes almost similar only
in the gapless state, when $1/(\tau_{m}\Delta)=1$. From these plots we can
conclude that, in contrast to elastic impurities, the ratio
$\mathcal{L}^{\mathrm{(sm)}}(E)/(\nu(E))^2$ increases with decreasing energy,
i.e., relative LDoS variations become stronger at smaller energy. We also found
that at the point where the mean-field average DoS vanishes, the correlation
function remains finite. Note again that the vanishing of the average DoS is
not exact result but only a consequence of the mean-field approximation. In
reality, the mean-field gap point marks the approximate location of transition
between delocalized quasiparticles and the Lifshitz-tail region corresponding
to localized quasiparticles.\cite{Simons01,ShytovPRL03} Calculation of the LDoS
correlation function is beyond the mean-field approach. As $\mathcal{L}(E)$
increases with decreasing the diffusion constant, the found growth of
$\mathcal{L}^{\mathrm{(sm)}}(E)$ with decreasing energy can be interpreted as
indication of slowing down diffusion when energy approaches the mobility edge. On
the other hand, our calculation is only applicable to delocalized diffusive
quasiparticles. One can expect that in the tail region the LDoS correlation
function should decay exponentially at the localization length. Therefore
measurements of the LDoS correlation function can be used to locate the
mobility edge separating delocalized and localized states.

\section{d-wave superconductors}

\begin{figure}[th]
\begin{center}
\includegraphics[width=2.0in]{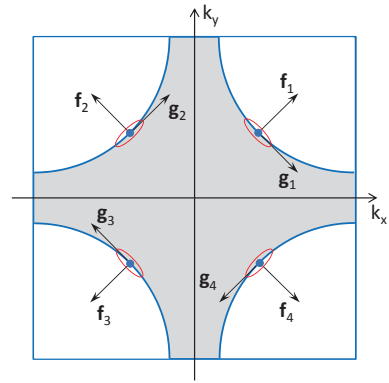}
\end{center}
\caption{Schematic Fermi surface for a cuprate d-wave superconductor. $\mathbf{f}_{n}$ and $\mathbf{g}_{n}$
represent, respectively, the unity vectors parallel to the Fermi velocity
$\mathbf{v}_{\mathrm{f}}$ and the ``gap velocity'' $\mathbf{v}_{\mathrm{g}}$ at
the n-th gap node. The ellipses illustrate the constant-energy surfaces near the nodes.} \label{dfs}
\end{figure}
Finally, let us consider the inhomogeneous LDoS for superconductors with d-wave
symmetry of the order parameter, which is of special interest because of its
relevance to cuprate high-temperature superconductors. Behavior of LDOS in
these materials was extensively studied by STM.\cite{STMcupr} Formally, the
long-range LDoS correlation function is still determined by the general Eq.\
(\ref{TwoCoopSC}) but the specifics of nodal gap structure has to be taken into
account.

We consider a two-dimensional d-wave superconductor with electronic spectrum
$\epsilon (k_x,k_y)$ defined within the square Brillouin zone
$|k_x|,|k_y|<\pi/a$ and characterized by the bandwidth $t\!=\!\epsilon
(\pi/a,0)-\epsilon (0,0)$, where $a$ is the lattice constant.
Figure \ref{dfs} illustrates the typical Fermi surface of a cuprate
superconductor \cite{dwaveFS} which, as usual, is determined by the equation
$\epsilon (k_x,k_y)=\mu$, where $\mu$ is the chemical potential. The order
parameter of the d$_{\mathrm{x^{2}-y^{2}}}$-wave pairing state is expressed by
\[
\Delta_{\mathbf{k}}=\Delta_{0}\left[  \cos (k_{x}a)-\cos (k_{y}a)\right].
\]
The gap nodes correspond to the points $\mathbf{k}_{n}=\pm\left(  k_{0}\pm
k_{0}\right)$ ($n=1,2,3,4$), where $k_{0}$ is determined by $\epsilon
(k_0,k_0)=\mu$. The quasiparticle spectrum is given by
$\varepsilon_{\mathbf{k}}\!=\sqrt{\xi_{\mathbf{k}}^{2}+\Delta_{\mathbf{k}}^{2}}$
 with $\xi_{\mathbf{k}}\!=\epsilon (k_x,k_y)\!-\!\mu$ and
close to the gap nodes it can be linearized as
$\varepsilon_{\mathbf{k}}\approx\sqrt{( v_{\mathrm{f}}\tilde{k}_{\mathrm{f}})^{2}+
(v_{\mathrm{g}}\tilde{k}_{\mathrm{g}})^{2}}$, where
$\tilde{\mathbf{k}}$ is the momentum measured from the node $\mathbf{k}_{n}$
and $v_{\mathrm{g}}\!\approx v_{\mathrm{f}}\Delta_{0}/t\! \ll \! v_{\mathrm{f}}$. The directions of the Fermi velocity $\mathbf{v}_{\mathrm{f}}$ and
``gap velocity'' $\mathbf{v}_{\mathrm{g}}$ are depicted in Fig. \ref{dfs}.

The weak-localization effects for d-wave superconductors were considered in
Refs.\ \onlinecite{YashenkinPRL01} and \onlinecite{YangPRB04} for arbitrarily
strong impurity scattering. Here we limit ourselves with the simplest
situation of weak isotropic scattering by non-magnetic elastic impurities. The
one-particle Green's function has the same matrix structure as general
Eq.\ (\ref{AvSupGreen}):
\begin{equation}
G_{\mathbf{k}}^{R,A}\left(  E\right)  =\frac{\left[  \tilde{\epsilon}\pm
i\gamma\right]  \tau_{0}+\Delta_{\mathbf{k}}\tau_{1}+\xi_{\mathbf{k}}\tau_{3}
}{\left[  \tilde{\epsilon}\pm i\gamma\right]  ^{2}-\varepsilon_{\mathbf{k}
}^{2}}.\label{gd}
\end{equation}
Here $\tilde{\epsilon}$ is the effective energy renormalized by scattering,
$\gamma=\gamma\left( \tilde{\epsilon}\right)  $ is the impurity-induced
relaxation rate. Both of them are self-consistently determined by the
self-energy part $\tilde{\epsilon}\pm i\gamma\left(  \tilde{\epsilon}\right)
=E-\Sigma_{0}^{\pm}\left(  \tilde{\epsilon}\right) $, with
\begin{equation}
\Sigma_{0}^{\pm}\left(  \tilde{\epsilon}\right)  =uN_{n}\int\int\frac{dk_{x}
}{2\pi}\frac{dk_{y}}{2\pi}\frac{\tilde{\epsilon}\pm i\gamma\left(
\tilde{\epsilon}\right)  }{\left(  \tilde{\epsilon}\pm i\gamma\left(
\tilde{\epsilon}\right)  \right)  ^{2}-\varepsilon_{\mathbf{k}}^{2}
},\label{Sigma}
\end{equation}
$N_{n}$ being the number of nodes (=4 in our case), and integration is limited
by the region near one node. Performing integration in Eq.\ (\ref{Sigma}) with
the above quasiparticle spectrum $\varepsilon_{\mathbf{k}}$ and separating the
imaginary part of $\Sigma_{0}^{\pm}\left(  \tilde{\epsilon}\right)  $ one
finds the transcendental equation for determination of the relaxation rate
$\gamma(\tilde{\epsilon})$:
\begin{equation}
\ln\frac{\Delta_{0}}{\sqrt{\tilde{\epsilon}^{2}+\gamma^{2}}}+\frac
{\tilde{\epsilon}}{\gamma}\arctan\frac{\tilde{\epsilon}}{\gamma}=\frac{2\pi
v_{\mathrm{f}}v_{\mathrm{g}}}{uN_{n}}.\label{imsigma}
\end{equation}
At zero energy it reproduces the known
result\cite{GorkovKalugZhETF85,PLeePRL93,DahmPRB05},
\[
\gamma(0)=\gamma_{0}=\Delta_{0}\exp\left(  -\frac{2\pi v_{\mathrm{f}}v_{\mathrm{g}}}{uN_{n}
}\right)  ,
\]
while at large energies
\[
\gamma\left(  \tilde{\epsilon}\gg\gamma_{0}\right)  \approx\frac{\pi}{2}
\frac{\tilde{\epsilon}}{\ln(\tilde{\epsilon}/\gamma_{0})+1}\text{ }.
\]

\begin{figure}[ptb]
\begin{center}
\includegraphics[width=3.4in]{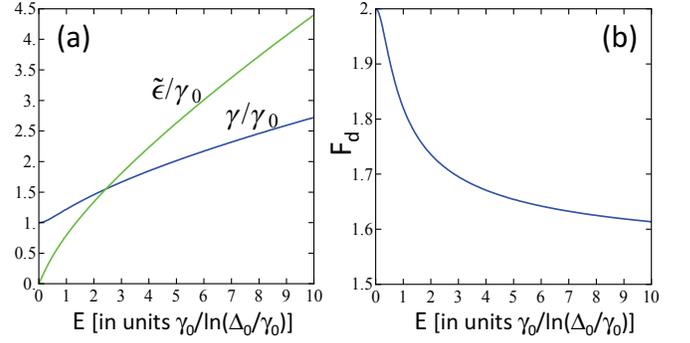}
\end{center}
\caption{(a)The energy dependences of the parameter $\tilde{\epsilon}$ and the
relaxation rate $\gamma$. (b) The energy dependence of the function $F_d$ which
determines the LDoS correlation function in Eq. (\ref{DoSCorrDWave1})}.
\label{dwavePlots}
\end{figure}

The real part of $\Sigma_{0}^{+}$ determines the value of $\tilde{\epsilon}$:
\begin{equation}
\tilde{\epsilon}-\frac{uN_{n}}{2\pi v_{\mathrm{f}}v_{\mathrm{g}}}\left(  \tilde{\epsilon}
\ln\frac{\Delta_{0}}{\sqrt{\gamma^{2}+\tilde{\epsilon}^{2}}}-\gamma
\arctan\frac{\tilde{\epsilon}}{\gamma}\right)  =E.\label{resigma}
\end{equation}
For small $E$, $\tilde{\epsilon}=\ln(\Delta_{0}/\gamma_{0})E$. When the energy
is large ($E\gg\gamma_{0})$
\[
\tilde{\epsilon}\approx E\frac{\ln\left(  \Delta_{0}/\gamma_{0}\right)  }
{\ln\left(  \tilde{\epsilon}/\gamma_{0}\right)  }.
\]
As follows from the structure of Eqs.\ (\ref{imsigma}) and (\ref{resigma}), the
energy dependences of  $\tilde{\epsilon}$ and $\gamma$ have scaling form,
$\tilde{\epsilon}=\gamma_0
\mathcal{G}_{\epsilon}[\ln(\Delta_0/\gamma_0)E/\gamma_0]$, $\gamma=\gamma_0
\mathcal{G}_{\gamma}[\ln(\Delta_0/\gamma_0)E/\gamma_0]$. These dependences are
presented in Fig.\ \ref{dwavePlots}(a)

The LDoS in superconducting state
is determined by the integral of the imaginary part of the Green's function
(\ref{gd}) which gives\cite{DurstLeePRB00}
\begin{equation}
N_{\mathrm{ex}}^{\left(  \mathrm{d}\right)  }(E)=\frac{N_{n}}{2\pi
^{2}v_{\mathrm{f}}v_{\mathrm{g}}}\left(  \gamma\ln\frac{\Delta_{0}^{2}}
{\tilde{\epsilon}^{2}+\gamma^{2}}+2\tilde{\epsilon}\arctan\frac{\tilde
{\epsilon}}{\gamma}\right)  .\label{dwaveDoS}
\end{equation}

The structure of cooperon for d-wave superconductors has been investigated in
Ref.\ \onlinecite{YashenkinPRL01,YangPRB04}. It was demonstrated that only
diagonal components of the cooperon $C^{\mathrm{(sd)}}_{kk}$ in the
Pauli-matrix expansion, Eq.\ (\ref{CooperonPauli}), are singular and these
singular components are connected by relation
$C^{\mathrm{(sd)}}_{00}=C^{\mathrm{(sd)}}_{11}=-C^{\mathrm{(sd)}}_{22}=C^{\mathrm{(sd)}}_{33}$.
Therefore the matrix
structure of d-wave cooperon is different from structure of the s-wave
cooperon, Eq.\ (\ref{SupCoop-s}). Derivation of the singular cooperon
$C_{\mathrm{d}}(q)\equiv C^{\mathrm{(sd)}}_{00}(q)$ for arbitrary energy is presented in
Appendix \ref{App-Cooperon-d} and result can be presented as
\begin{equation}
C_{\mathrm{d}}(q)=\frac{\gamma^{2}}{\pi N_{\mathrm{ex}}^{\left(
\mathrm{d}\right)  }\mathcal{D}q^{2}}\label{Coop-d1}
\end{equation}
where parameters $\gamma$, $N_{\mathrm{ex}}^{\left(  \mathrm{d}\right)  }$,
and the diffusion coefficient $\mathcal{D}$ are energy dependent with
\begin{equation}
\mathcal{D}\left(  E\right)  =\frac{\langle v^{2}\rangle}{2\gamma}\frac{1+\frac
{\tilde{\epsilon}}{\gamma}\arctan\frac{\tilde{\epsilon}}{\gamma}}{\ln
\frac{\Delta_{0}}{\sqrt{\tilde{\epsilon}^{2}+\gamma^{2}}}+\frac{\tilde
{\epsilon}}{\gamma}\arctan\frac{\tilde{\epsilon}}{\gamma}}.
\end{equation}
Note that our result for $C_{\mathrm{d}}(q)$ is smaller by factor four than the
result of Refs.\ \onlinecite{YashenkinPRL01,YangPRB04}. The origin of this
discrepancy is discussed in Appendix \ref{App-Cooperon-d}. For more
transparent representation of the energy dependence in Eq.\ (\ref{Coop-d1}),
we note a useful relation
\[
N_{\mathrm{ex}}^{\left(  \mathrm{d}\right)  }(E)\mathcal{D}(E)=N_{\mathrm{ex}}^{\left(
\mathrm{d}\right)  }(0)\mathcal{D}(0)[1+(\tilde{\epsilon}/\gamma)\arctan(\tilde
{\epsilon}/\gamma)].
\]

Calculation of the Hikami boxes $\mathcal{B}_{inj}^{\left(  \mathrm{d}\right)}$
defined by Eq.\ (\ref{Hikami-swave}) with Green's functions (\ref{gd}) (see
Appendix \ref{App-Hikami-d}), results in:
\begin{align*}
\mathcal{B}_{000}^{\left(  \mathrm{d}\right)  } &  =\frac{N_{n}}{v_{\mathrm{f}
}v_{\mathrm{g}}}\left[  \tilde{\epsilon}+i\gamma\right]  ^{2}\left[
\tilde{\epsilon}-i\gamma\right]  J^{(0)},\\
\mathcal{B}_{101}^{\left(  \mathrm{d}\right)  } &  =\mathcal{B}_{303}^{\left(
\mathrm{d}\right)  }=\frac{N_{n}}{v_{\mathrm{f}}v_{\mathrm{g}}}\left[
\tilde{\epsilon}-i\gamma\right]  J^{(2)},\\
\mathcal{B}_{330}^{\left(  \mathrm{d}\right)  } &  =\mathcal{B}_{033}^{\left(
\mathrm{d}\right)  }=\mathcal{B}_{011}^{\left(  \mathrm{d}\right)
}=\mathcal{B}_{110}^{\left(  \mathrm{d}\right)  }=\frac{N_{n}}{v_{\mathrm{f}
}v_{\mathrm{g}}}\left[  \tilde{\epsilon}+i\gamma\right]  J^{(2)},
\end{align*}
where
\begin{align*}
J^{(0)} &  =-\frac{i}{16\pi\left(  \tilde{\epsilon}+i\gamma\right)  ^{2}
\tilde{\epsilon}\gamma}\left[  1+\frac{\left(  \tilde{\epsilon}+i\gamma
\right)  ^{2}}{\tilde{\epsilon}\gamma}\arctan\frac{\tilde{\epsilon}}{\gamma
}\right]  ,\\
J^{(2)} &  =-\frac{i}{32\pi\tilde{\epsilon}\gamma}\left[  1+\frac{\left(
\tilde{\epsilon}-i\gamma\right)  ^{2}}{\tilde{\epsilon}\gamma}\arctan
\frac{\tilde{\epsilon}}{\gamma}\right]  .
\end{align*}
Further summation of these, non-zero, components of $\mathcal{B}
_{inj}^{\left(  \mathrm{d}\right)  }$ in Eq.\ (\ref{VertexSC}) with
corresponding traces of five Pauli-matrices\ products as the coefficients
allows to present the tensor $U_{km}$ in the diagonal form $U_{km}^{\left(
\mathrm{d}\right)  }=U_{k}\delta_{km}$ with
\begin{align*}
U_{0;2} &  =\mathcal{B}_{000}^{\left(  \mathrm{d}\right)  }+2\mathcal{B}
_{101}^{\left(  \mathrm{d}\right)  }\pm4\mathcal{B}_{011}^{\left(
\mathrm{d}\right)  }\\
U_{1;3} &  =\mathcal{B}_{000}^{\left(  \mathrm{d}\right)  }+2\mathcal{B}
_{101}^{\left(  \mathrm{d}\right)  }.
\end{align*}
In result, the LDoS correlation function Eq.\ (\ref{TwoCoopSC}) for the d-wave
case takes the form
\begin{align}
&  \mathcal{L}_{\mathrm{ex}}^{\mathrm{(d)}}(\mathbf{r}\!-\!\mathbf{r}^{\prime}
,E)\!=\!\frac{8}{\pi^{2}}\left[  \left\vert \mathcal{B}_{000}^{\left(
\mathrm{d}\right)  }\!+\!2\mathcal{B}_{101}^{\left(  \mathrm{d}\right)
}\right\vert ^{2}\!+\!8|\mathcal{B}_{011}^{\left(  \mathrm{d}\right)  }
|^{2}\right]  C_{\mathrm{d}}^{2}(\mathbf{r}\!-\!\mathbf{r}^{\prime
})\nonumber\\
&  =\frac{[N_{n}^2\gamma\left(  \tilde{\epsilon}\right)  ]^{2}}{8\pi^{4}
v_{\mathrm{f}}^{2}v_{\mathrm{g}}^{2}\gamma_{0}^{4}}F_{\mathrm{d}}\left(  \frac
{\tilde{\epsilon}}{\gamma\left(  \tilde{\epsilon}\right)  }\right)
C_{\mathrm{d}}^{2}(\mathbf{r}\!-\!\mathbf{r}^{\prime}
,E=0),\label{DoSCorrDwave}
\end{align}
with $\gamma\left(  \tilde{\epsilon}\right)  $ and $\tilde{\epsilon}(E)$ have
to be determined from Eqs.\ (\ref{imsigma}) and (\ref{resigma}),
\begin{align*}
F_{\mathrm{d}}(x) &  =\frac{1+1/x^{2}}{\left(  1+x\arctan
x\right)  ^{2}}\left\{  \frac{3}{2}\left[  1+\frac{x^{2}-1}{x}\arctan
x\right]  ^{2}\right. \\
&  \left.  +2\arctan^{2}x\right\}  =\left\{
\begin{tabular}
[c]{ll}
$2$ & for $x=0$\\
$3/2$ & for $x\gg1$
\end{tabular}
\ \ \right.  ,
\end{align*}
and the zero-energy cooperon in real space is given by
\[
C_{\mathrm{d}}(r,E=0)=\frac{4\pi}{N_{n}}\frac{v_{\mathrm{f}}v_{\mathrm{g}}\gamma_{0}^{2}
}{\left\langle v^{2}\right\rangle }\ln\frac{l_{\phi}}{r}.
\]
Using this result, we can rewrite LDoS correlation function in a more
transparent form
\begin{align}
&  \mathcal{L}^{\mathrm{(d)}}_{\mathrm{ex}}(\mathbf{r}\!-\!\mathbf{r}^{\prime}
,E)=[N_{\mathrm{ex}}^{\left(  \mathrm{d}\right)  }(E)]^{2}F_{\mathrm{d}}\left(
\frac{\tilde{\epsilon}}{\gamma}\right) \nonumber\\
&  \times\frac{u^{2}}{32\pi^{2}\left\langle v^{2}\right\rangle ^{2}
}\ln^{2}\frac{l_{\phi}}{|\mathbf{r}-\mathbf{r}^{\prime}|}\label{DoSCorrDWave1}
\end{align}
We can see that, in contrast to s-wave superconductors with potential
impurities, the energy dependence of the correlation function is not determined
by the average density of states. Additional dependence characterized by the
function $F_{\mathrm{d}}[\tilde{\epsilon(E)}/\gamma(E)]$ appears. Both the
ratio $\tilde{\epsilon}/\gamma$ and function $F_d$ have universal dependences
on the scaled energy $(E/\gamma_0)\ln(\Delta_{0}/\gamma_0)$. The energy
dependence of $F_d$ is plotted in Fig.\ \ref{dwavePlots}(b). We observe the
same tendency as for an s-wave superconductor with magnetic impurities: the
relative LDoS variations become stronger at smaller energies.

\section{Discussion}

It is known that observable quantities of disordered metallic systems exhibit
large mesoscopic fluctuations.\cite{Mesosc} The large-scale correlations of the
LDoS demonstrates long tails extending up to $r\sim l_{\phi}$.\cite{AKL} As we
have seen above, due to diffusive propagation of quasiparticles, such property
of the LDoS correlation function remains valid also in superconductive state.
This correlation function for all cases depends on scatterers potential $V$ in
the same way: $\mathcal{L} \sim V^4$ and its spatial dependence is determined
only by the dimensionality of superconductor. Our analysis demonstrates,
however, that the \emph{energy dependence} of the large-scale LDoS correlations
carries valuable information about the order parameter symmetry and character
of scattering in various superconductive systems. Indeed, as we have seen
above, while the spatial dependence of
$\mathcal{L}(\mathbf{r}\!-\!\mathbf{r}^{\prime},E)$ is the same for all types
of considered superconducting systems and is determined by the square of
normal-metal cooperon, the energy dependence of the magnitude of such
correlator can serve as the fingerprint of superconductor intrinsic properties.
In the reference case of s-wave superconductor with elastic impurities such
energy dependence is just given by the square of the BCS quasi-particle density
of states. More complex symmetry of the order parameter manifests itself in
significant changes of the energy dependence of LDoS correlation function.
Investigated above the case of d-wave pairing (see Fig.\ \ref{dwavePlots}(b)),
demonstrated that the corresponding LDoS correlation function, normalized on
the appropriate square of the LDoS, already depends on the quasi-particle
energy, although this dependence is rather smooth.

To illustrate the role of pair-breaking scattering, we analyzed the case of s-wave
superconductor containing magnetic impurities. It is worth to discuss our
results in light of the extended AG theory in Ref.\ \onlinecite{Simons01}, which
investigated the low-energy behavior of the average DoS in the framework of the
nonlinear sigma model. The authors of Ref.\ \onlinecite{Simons01} demonstrate,
that AG results, establishing the formation of the hard gap in the local
density of states, correspond to the saddle-point solution of the proposed
effective action. Their non-perturbative extension beyond the mean-field
approximation results in appearance of sub-gap exponential tails in the density
of states, what should lead also to appearance of nonzero moments of this
physical value for energies below the AG gap. Plausibly, our perturbative
calculus of the second moment of local densities of states
$\mathcal{L}^{(\mathrm{sm})}(\mathbf{r}-\mathbf{r}^{\prime},E)$ (see Fig.
\ref{RLDoSMag}) indicates the same phenomenon. %
Our analysis allows for a straightforward generalization to other similar
situations, such as interband scattering in multiband superconductors with
different signs of order parameter in different bands ($s_{\pm}$ state), a
widely discussed model for iron pnictides and selenides.

Modern STM technique, in principle, allows to probe the long-range correlations
studied theoretically in this article. The challenge is to perform scan over
large areas with sizes exceeding mean-free path. Another interesting
opportunity provided by STM measurements of the LDoS correlation function is to
extract and  study the temperature-depending phase-breaking length
$l_{\phi}(T)$ of quasiparticle excitations in superconductors of different
nature. As far as we know, such a problem was never considered neither
theoretically nor experimentally.

\begin{acknowledgments}
The authors acknowledge valuable discussions with V.~E.~Kravtsov,
K.~A.~Matveev, and I.~E.~Smolyarenko. This work was supported by UChicago
Argonne, LLC, operator of Argonne National Laboratory, a U.S. Department of
Energy Office of Science laboratory, operated under contract No.
DE-AC02-06CH11357. A.A.V. acknowledges support of the MIUR under the project
PRIN 2008 and the European Community FP7-IRSES programs: ``ROBOCON'' and
``SIMTECH''.
\end{acknowledgments}

\appendix
\section{ s-wave superconductor with elastic impurities}

\subsection{Calculation of s-wave cooperon \label{App-Cooperon-s}}

\vspace{-0.1in}
\begin{figure}[h]
\begin{center}
\includegraphics[ width=3.4in ]{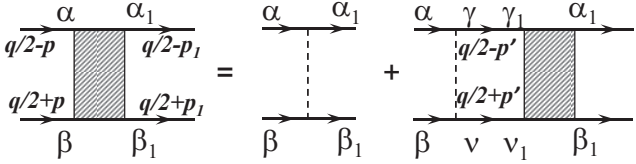}
\end{center}
\caption{Graphic presentation of Eq.\ (\ref{EqSCcoop}) for the superconducting
cooperon. }
\label{Fig-PauliVertex}
\end{figure}
\vspace{-0.1in}
In this Appendix we consider the cooperon structure for a
s-wave superconductor with weak potential scattering. The superconducting
cooperon has four Nambu indices and obeys the equation graphically represented
Fig.\ \ref{Fig-PauliVertex},
\begin{align}
C_{\beta\beta_{1}}^{\alpha\alpha_{1}}  &  =u_{\beta\beta_{1}}^{\alpha
\alpha_{1}}+ u_{\beta\nu}^{\alpha\gamma}\Pi_{\nu\nu_{1}}^{\gamma\gamma_{1}
}C_{\nu_{1}\beta_{1}}^{\gamma_{1}\alpha_{1}}\label{EqSCcoop}\\
\Pi_{\nu\nu_{1}}^{\gamma\gamma_{1}}  &  =\int\frac{d^{D}\mathbf{p}
}{(2\pi)^{D}}\hat{G}_{\gamma\gamma_{1}}^{A}(\mathbf{p})\hat{G}_{\nu\nu_{1}
}^{R}(\mathbf{p}-\mathbf{q})\nonumber
\end{align}
where we again assume summation with respect to repeated Nambu indices. Here
the superconducting Green's functions $\hat{G}^{R,A}$ averaged over impurities
are given by Eq.\ (\ref{AvSupGreen}). For Born potential impurities, the matrix
impurity line is given by
\[
u_{\beta\beta_{1}}^{\alpha\alpha_{1}}=u\hat{\tau}_{\alpha\alpha_{1}}^{3}
\hat{\tau}_{\beta\beta_{1}}^{3}\text{, }
\]
with $u=n_{i}V^{2}=1/(2\pi\nu\tau)$.

To proceed, we represent the matrix impurity line as
\begin{align}
u_{\beta\beta_{1}}^{\alpha\alpha_{1}}  &  =\sum_{i,j=0}^{3}u_{ij}\hat{\tau
}_{\alpha\beta}^{i}\hat{\tau}_{\beta_{1}\alpha_{1}}^{j}\label{ImpLineDecomp}\\
\text{with }u_{ij}  &  =\frac{u}{2}E_{ji}^{(33)},\nonumber
\end{align}
where, following Refs. \onlinecite{Kulik,VarlamovDorinZhETF86}, we introduced
notation for the trace of four Pauli matrices
\[
E_{lm}^{(ij)}\equiv\frac{1}{2}\mathrm{Tr}\left(  \hat{\tau}^{i}\hat{\tau}
^{l}\hat{\tau}^{j}\hat{\tau}^{m}\right)  \text{.}
\]
We will need only the following components of $E_{ml}^{(ij)}$ which have a simple
block structure \cite{Kulik,VarlamovDorinZhETF86}
\begin{align*}
\hat{E}^{(00)}  &  =
\begin{bmatrix}
\hat{\tau}^{0} & 0\\
0 & \hat{\tau}^{0}
\end{bmatrix}
,\ \hat{E}^{(11)}=
\begin{bmatrix}
\hat{\tau}^{0} & 0\\
0 & -\hat{\tau}^{0}
\end{bmatrix}
,\ \hat{E}^{(33)}=
\begin{bmatrix}
\hat{\tau}^{3} & 0\\
0 & -\hat{\tau}^{3}
\end{bmatrix}
\\
\hat{E}^{(01)}  &  =
\begin{bmatrix}
\hat{\tau}^{1} & 0\\
0 & \hat{\tau}^{2}
\end{bmatrix}
,\ \hat{E}^{(10)}=
\begin{bmatrix}
\hat{\tau}^{1} & 0\\
0 & -\hat{\tau}^{2}
\end{bmatrix}
\end{align*}
Since $E_{ij}^{(33)}$ is diagonal, we have $u_{ij}=\frac{u}{2}\alpha_{i}
\delta_{ij}$ with $\alpha_{0}=\alpha_{3}=-\alpha_{1}=-\alpha_{2}=1$.

We use a similar decomposition for the polarization operator
\begin{subequations}
\begin{equation}
\Pi_{\nu\nu_{1}}^{\gamma\gamma_{1}}(\mathbf{q},E)=\nu \sum_{l,m=0}^{3}S_{lm}
\hat{\tau}_{\gamma\nu}^{l}\hat{\tau}_{\gamma_{1}\nu_{1}}^{m}
\label{sPolFunPres}
\end{equation}
with
\begin{align}
S_{lm}  &  =\frac{1}{2\nu}\mathrm{Tr}\left[  \int\frac{d^{D}\mathbf{p}}{(2\pi
)^{D}}\hat{G}_{\mathbf{p}}^{A}\hat{\tau}_{l}\hat{G}_{\mathbf{p}-\mathbf{q}
}^{R}\hat{\tau}_{m}\right] \nonumber\\
&  =E_{lm}^{(ij)}P_{ij}\label{Slm}
\end{align}
and
\begin{equation}
P_{ij}(\mathbf{q})=\left\langle \int d\xi_{\mathbf{p}}G_{i,\mathbf{p}}
^{A}G_{j,\mathbf{p}-\mathbf{q}}^{R}\right\rangle _{F},\label{Pij}
\end{equation}
where $\left\langle \ldots\right\rangle _{F}$ means averaging over the Fermi
surface. Nonzero components of $P_{ij}$ are only for subscripts $(i,j)=(0,0)$,
$(1,1)$, $(0,1)$, $(1,0)$, $(3,3)$ and\textbf{\ }they can be straightforwardly
calculated as
\end{subequations}
\begin{align*}
P_{00}  &  =\pi\tau\gamma_{\mathbf{q}}\frac{E^{2}}{E^{2}-\Delta^{2}}
,\ P_{11}=\pi\tau\gamma_{\mathbf{q}}\frac{\Delta^{2}}{E^{2}-\Delta^{2}}\\
P_{01}  &  =P_{10}=\pi\tau\gamma_{\mathbf{q}}\frac{E\Delta}{E^{2}-\Delta^{2}
},\ P_{33}=\pi\tau\gamma_{\mathbf{q}}
\end{align*}
with
\begin{equation}
\gamma_{\mathbf{q}}\equiv\left\langle \frac{1}{1+\tau^{2}\left(
\mathbf{v}_{F}\mathbf{q}\right)  ^{2}}\right\rangle \approx1-\tau
^{2}\left\langle \left(  \mathbf{v}_{F}\mathbf{q}\right)  ^{2}\right\rangle
.\label{difpole}
\end{equation}
Due to the block structure of the corresponding components of $E_{lm}^{(ij)}$,
the $4\times4$ matrix $\hat{S}$ is composed of two independent $2\times2$
submatrices, $[0,1]$ and $[2,3]$ blocks
\begin{equation}
\hat{S}=
\begin{bmatrix}
\hat{S}_{A} & 0\\
0 & \hat{S}_{B}
\end{bmatrix}
.\label{Sblocks}
\end{equation}
These blocks can be explicitly found
\begin{align*}
\hat{S}_{A}  &  =\hat{\tau}^{0}\left(  P_{00}+P_{11}\right)  +\hat{\tau}
^{3}P_{33}+2\hat{\tau}^{1}P_{01},\\
\hat{S}_{B}  &  =\hat{\tau}^{0}\left(  P_{00}-P_{11}\right)  -\hat{\tau}
^{3}P_{33}.
\end{align*}
Note that, due to the relation $P_{00}-P_{11}=P_{33}$, the component $S_{22}$ vanishes.

Equation for the Pauli-matrix components of the cooperon $C^{\left(  \mathrm{s}\right)}_{ij}=\frac{1}
{4}\hat{\tau}_{\beta\alpha}^{i}C_{\beta\beta_{1}}^{\alpha\alpha_{1}}\hat{\tau
}_{\beta_{1}\alpha_{1}}^{j}$ now takes the form
\[
C^{\left(  \mathrm{s}\right)}_{ij}=\frac{u}{2}\alpha_{i}\delta_{ij}+\frac{1}{2\pi\tau}\alpha_{i}
S_{il}C^{\left(  \mathrm{s}\right)}_{lj}
\]
The matrix $\hat{C}{\left(  \mathrm{s}\right)}$ has the same block form as $\hat{S}$ meaning that this
$4\times4$ system splits into two independent $2\times2$ subsystems which
allows us to obtain analytical results
\begin{align}
C^{\left(  \mathrm{s}\right)}_{A}  &  =\frac{u}{2}\left[  1-\gamma_{\mathbf{q}}\right]  ^{-1}\nonumber\\
&  \times\left\{  \left(  1-\frac{\gamma_{\mathbf{q}}}{2}\right)  \hat{\tau
}_{3}+\frac{\gamma_{\mathbf{q}}}{2}\frac{E^{2}+\Delta^{2}}{E^{2}-\Delta^{2}
}\hat{\tau}_{0}-\gamma_{\mathbf{q}}\frac{E\Delta}{E^{2}-\Delta^{2}}\hat{\tau
}_{1}\right\}  ,\label{CscA}\\
C^{\left(  \mathrm{s}\right)}_{B}  &  =\frac{u}{2}
\begin{Bmatrix}
-1 & 0\\
0 & \left(  1-\gamma_{\mathbf{q}}\right)  ^{-1}
\end{Bmatrix}
.\label{CscB}
\end{align}
We can see that $C^{\left(  \mathrm{s}\right)}_{33}(\mathbf{q})$ equals half of the normal-state
cooperon,
\begin{equation}
C^{\left(  \mathrm{s}\right)}_{33}(\mathbf{q})=\frac{1}{2}C_{\mathrm{n}}(\mathbf{q})=\frac{1}{4\pi\nu\tau}\frac
{\mathrm{D}}{l^{2}q^{2}}\label{C33}
\end{equation}
The singular part of $C^{\left(  \mathrm{s}\right)}_{A}$ for $q\rightarrow0$ is
\begin{align}
C^{\left(  \mathrm{s}\right)}_{A}  &  =\frac{u/2}{\tau^{2}\left\langle \left(  \mathbf{v}_{F}
\mathbf{q}\right)  ^{2}\right\rangle }\left\{  \frac{1}{2}\hat{\tau}_{3}
+\frac{1}{2}\frac{E^{2}+\Delta^{2}}{E^{2}-\Delta^{2}}\hat{\tau}_{0}
-\frac{E\Delta}{E^{2}-\Delta^{2}}\hat{\tau}_{1}\right\} \nonumber\\
&  =\frac{u/2}{\tau^{2}\left\langle \left(  \mathbf{v}_{F}\mathbf{q}\right)
^{2}\right\rangle }
\begin{Bmatrix}
\frac{E^{2}}{E^{2}-\Delta^{2}} & -\frac{E\Delta}{E^{2}-\Delta^{2}}\\
-\frac{E\Delta}{E^{2}-\Delta^{2}} & \frac{\Delta^{2}}{E^{2}-\Delta^{2}}
\end{Bmatrix}
.\label{CscA-sing}
\end{align}
This allows us to present the cooperon components in the following
form\emph{\ }
\[
C^{\left(  \mathrm{s}\right)}_{ij}=C^{\left(  \mathrm{s}\right)}_{00}(-\Delta/E)^{i+j},\qquad\left(  i,j=0,1\right)
\]
with
\[
C^{\left(  \mathrm{s}\right)}_{00}=\frac{E^{2}}{E^{2}-\Delta^{2}}C^{\left(  \mathrm{s}\right)}_{33}.
\]
Note that in the normal-state limit, $\Delta\rightarrow0$, only the components
$C^{\left(  \mathrm{s}\right)}_{00}$ and $C^{\left(  \mathrm{s}\right)}_{33}$ remain singular.

\subsection{Calculation of s-wave Hikami boxes \label{App-Hikami-s}}

In this Appendix we present details of calculations of Hikami boxes
$\mathcal{B}_{inj}^{\left(  \mathrm{s}\right)  }$ defined by
Eq.\ (\ref{Hikami-swave}). In $\mathbf{k}$-space $\mathcal{B}_{inj}^{\left(
\mathrm{s}\right)  }$ are given by the integrals
\begin{align}
\mathcal{B}_{inj}^{\left(  \mathrm{s}\right)  } &  =\int\frac{d^{D}\mathbf{k}
}{(2\pi)^{D}}g_{i}^{A}(\epsilon_{\mathbf{k}})g_{n}^{R}(\epsilon_{\mathbf{k}
})g_{j}^{A}(\epsilon_{\mathbf{k}})\label{HikamiSP}\\
&  =\nu\int d\xi g_{i}^{A}(\xi)g_{n}^{R}(\xi)g_{j}^{A}(\xi)\nonumber
\end{align}
where $g_{i}^{R,A}$ are defined in Eq.(\ref{GreenPauli}).

For example, the component $\mathcal{B}_{000}^{\left(  \mathrm{s}\right)  }$ is
given by the integral
\[
\mathcal{B}_{000}^{\left(  \mathrm{s}\right)  }\!=\!\nu\!\int\!d\xi
\frac{\left(  \alpha^{A}\right)  ^{2}\alpha^{R}E^{3}}{\left[  \left(
\alpha^{A}\right)  ^{2}\left(  E^{2}\!-\!\Delta^{2}\right)  \!-\!\xi
^{2}\right]  ^{2}\!\left[  \left(  \alpha^{R}\right)  ^{2}\left(
E^{2}\!-\!\Delta^{2}\right)  \!-\!\xi^{2}\right]  }
\]
Performing integration and substituting $\alpha^{R,A}$, see
Eq.\ (\ref{AvSupGreen}), we obtain
\[
\mathcal{B}_{000}^{\left(  \mathrm{s}\right)  }=-\frac{i\pi\nu\tau^{2}}
{2}\frac{E^{3}\left(  \sqrt{E^{2}-\Delta^{2}}+3i/2\tau\right)  }{\left(
E^{2}-\Delta^{2}\right)  ^{3/2}\left(  \sqrt{E^{2}-\Delta^{2}}+i/2\tau\right)
}
\]
As $g_{1}^{R,A}=(\Delta/E)g_{0}^{R,A}$, the components $\mathcal{B}
_{imj}^{\left(  \mathrm{s}\right)  }$ for $i,m,j=0,1$ are connected with the
$\mathcal{B}_{000}$ by the simple relation
\[
\mathcal{B}_{imj}^{\left(  \mathrm{s}\right)  }=\mathcal{B}_{000}^{\left(
\mathrm{s}\right)  }(\Delta/E)^{i+j+m}.
\]
Two remaining nonzero components, $\mathcal{B}_{033}^{\left(  \mathrm{s}
\right)  }$ and $\mathcal{B}_{303}^{\left(  \mathrm{s}\right)  }$, are given by
\begin{widetext}
\vspace{-0.2in}
\begin{equation*}
\dbinom{\mathcal{B}_{033}^{\left( \mathrm{s}\right) }}{\mathcal{B}%
_{303}^{\left( \mathrm{s}\right) }}=-\nu \dbinom{\alpha ^{A}}{\alpha ^{R}}E%
\int d\xi \frac{\xi ^{2}}{\left[ \xi ^{2}-\left( \sqrt{E^{2}-\Delta ^{2}}+%
\frac{i}{2\tau }\right) ^{2}\right] ^{2}\left[ \xi ^{2}-\left( \sqrt{%
E^{2}-\Delta ^{2}}-\frac{i}{2\tau }\right) ^{2}\right] }
\end{equation*}
\vspace{-0.15in}
\end{widetext}
and evaluation of the integral gives
\begin{align*}
\mathcal{B}_{033}^{\left(  \mathrm{s}\right)  } &  =-\frac{i\pi\nu E\tau^{2}
}{2\sqrt{E^{2}-\Delta^{2}}},\\
\mathcal{B}_{303}^{\left(  \mathrm{s}\right)  } &  =\frac{\alpha^{R}}
{\alpha^{A}}\mathcal{B}_{033}^{\left(  \mathrm{s}\right)  }=-\frac{i\pi\nu
E\tau^{2}}{2\sqrt{E^{2}-\Delta^{2}}}\frac{\sqrt{E^{2}-\Delta^{2}}-\frac
{i}{2\tau}}{\sqrt{E^{2}-\Delta^{2}}+\frac{i}{2\tau}}.
\end{align*}

\section{s-wave superconductor with magnetic impurities}

\subsection{Calculation of cooperon\label{App-Cooperon-sm}}

In this appendix we present computation details of the cooperon for s-wave
superconductor with Ising magnetic impurities polarized along $z$ axis.
The equation for the cooperon graphically presented in Fig. \ref{Fig-PauliVertex}
is again given by Eq.\ (\ref{EqSCcoop}) but the impurity line here is now
composed by potential and magnetic contributions
\[
u_{\beta\beta_{1}}^{\alpha\alpha_{1}}=u\hat{\tau}_{\alpha\alpha_{1}}^{3}
\hat{\tau}_{\beta_{1}\beta}^{3}+u_{m}\delta_{\alpha\alpha_{1}}\delta
_{\beta\beta_{1}}
\]
with $u_{m}=1/(2\pi\nu\tau_{m})$. In order to obtain the explicit expression
for the cooperon we follow the same route as in the case of potential
impurities. We again present the impurity line in the form
(\ref{ImpLineDecomp}), where the $4\times4$ matrix $u_{ij}$ is now given by
\[
u_{ij}=\frac{u_{i}}{2}\delta_{ij},\ u_{i}=uE_{ii}^{(33)}+u_{m}.
\]
Using this presentation and formulas for the polarization function (see Eqs.
(\ref{sPolFunPres}) and (\ref{Slm})) one can write the equation for the
cooperon:
\begin{equation}
C_{ij}^{\mathrm{(sm)}}=\frac{u_{i}}{2}\delta_{ij}+\nu u_{i}S_{il}C_{lj}^{\mathrm{(sm)}}.\label{coopmag}
\end{equation}
To evaluate the matrix $S_{il}$, we again have to compute the integrals
$P_{ij}$ defined by Eq.\ (\ref{Pij}) but accounting for scattering from
magnetic impurities. For nonzero components, we obtain
\begin{align*}
P_{00}  &  =\pi\tau_{\mathrm{qp}}\tilde{\gamma}_{\mathbf{q}}\frac{|\eta_{+}|^{2}}{\left\vert
\eta_{+}^{2}-1\right\vert },\ P_{11}=\pi\tau_{\mathrm{qp}}\tilde{\gamma}_{\mathbf{q}}\frac
{1}{\left\vert \eta_{+}^{2}-1\right\vert },\\
P_{01}  &  =\pi\tau_{\mathrm{qp}}\tilde{\gamma}_{\mathbf{q}}\frac{\eta_{+}}{\left\vert \eta
_{+}^{2}-1\right\vert },\ P_{33}\approx\pi\tau_{\mathrm{qp}}\tilde{\gamma}_{\mathbf{q}}
\end{align*}
where $\eta_{+}$ is introduced by Eq.\ (\ref{SelfConsuRA}) and
$\tilde{\gamma}_{\mathbf{q}} $ can be obtained from  $\gamma_{\mathbf{q}}$
defined in Eq.\ (\ref{difpole}) by the replacement of $\tau $ with the
energy-dependent quasiparticle relaxation time
$\tau_{\mathrm{qp}}=1/(2\operatorname{Im}\sqrt{\tilde{E}_{+}^{2}-\Delta
_{+}^{2}})$. The useful expression for the latter parameter is derived in
Appendix \ref{App-QPtime}
\begin{align*}
1/\tau_{\mathrm{qp}}
=\frac{1}{\tau}+\frac{\chi_1}{\tau_{m}}\ \ \
\text{with }\chi_1=\frac{|\eta_{+}
|^{2}+1}{\left\vert \eta_{+}^{2}-1\right\vert }.
\end{align*}

Equation\ (\ref{coopmag}) for the cooperon again splits into two independent
$2\times2$ blocks
\begin{align*}
\hat{C}_{A}^{\mathrm{(sm)}}  &  =\frac{1}{2}\hat{u}^{A}+2\nu\hat{u}^{A}\hat{S}^{A}\hat{C}
_{A}^{\mathrm{(sm)}},\\
\hat{C}_{B}^{\mathrm{(sm)}}  &  =\frac{1}{2}\hat{u}^{B}+2\nu\hat{u}^{B}\hat{S}^{B}\hat{C}_{B}^{\mathrm{(sm)}},
\end{align*}
where the blocks of the matrices $\hat{u}$ and $\hat{S}$ can be evaluated as
$\hat{u}^{A}=\left(  u/2\right)  \hat{\tau}^{3}+\left(  u_{m}/2\right)
\hat{\tau}^{0}$, $\hat{u}^{B}=-\left(  u/2\right)  \hat{\tau}^{3}+\left(
u_{m}/2\right)  \hat{\tau}^{0}$, and
\begin{align*}
\hat{S}^{A}  &  =\hat{\tau}^{0}\left(  P_{00}+P_{11}\right)  +\hat{\tau}
^{3}P_{33}+2\hat{\tau}^{1}\operatorname{Re}P_{01},\\
\hat{S}^{B}  &  =\hat{\tau}^{0}\left(  P_{00}-P_{11}\right)  -\hat{\tau}
^{3}P_{33}+2i\hat{\tau}^{2}\operatorname{Im}P_{01}.
\end{align*}
Deriving relation
\begin{align*}
\nu\hat{u}^{A}\hat{S}^{A}  &  =\tau_{\mathrm{qp}}\tilde{\gamma}_{\mathbf{q}} \left[ \hat{\tau
}^{0}\left(  \frac{1}{2\tau}+\frac{\chi_1}{2\tau_{m}}\right)
  \left.  +\hat{\tau}^{3}\left(  \frac{\chi_1}{2\tau}+\frac{1}{2\tau_{m}}\right)\right.  \right.
\\
&  \left.  +\left(  \frac{i}{\tau}\hat{\tau}^{2}+\frac{1}{\tau_{m}}\hat{\tau
}^{1}\right)  \frac{\operatorname{Re}\eta_{+}}{\left\vert \eta_{+}^{2}
-1\right\vert }\right],
\end{align*}
we can formally present $\hat{C}^{\mathrm{(sm)}}_{A}$ as
\begin{align*}
\hat{C}^{\mathrm{(sm)}}_{A}  &  =\left\{\hat{\tau}_{0}\left[  1\!-\!\tau_{\mathrm{qp}}\tilde{\gamma}_{\mathbf{q}}
\left(\frac{1}{2\tau}\!+\!\frac{\chi_1}{2\tau_{m}}\right)\right]
\!-\!\hat{\tau}_{3}\tau_{\mathrm{qp}}\tilde{\gamma}_{\mathbf{q}}\left[  \frac{\chi_1}{2\tau}\!+\!\frac{1}{2\tau_{m}
}\right] \right. \\
&  \left.  -\tau_{\mathrm{qp}}\tilde{\gamma}_{\mathbf{q}}\frac{\operatorname{Re}\eta_{+}}{\left\vert
\eta_{+}^{2}-1\right\vert }\left[  \frac{i}{\tau}\hat{\tau}^{2}+\frac{1}
{\tau_{m}}\hat{\tau}^{1}\right]  \right\}  ^{-1}\frac{1}{2}\left(  u\hat{\tau
}^{3}+u_{m}\hat{\tau}^{0}\right)  .
\end{align*}
After some algebra this expression can be transformed to much simpler form
\begin{align}
\hat{C}^{\mathrm{(sm)}}_{A}  &  =\frac{1/4\pi\nu}{1-\tilde{\gamma}_{\mathbf{q}}}\left\{  \left[  \frac{1}
{\tau_{m}}+\frac{\tau_{\mathrm{qp}}\tilde{\gamma}_{\mathbf{q}}\chi_1}{2}\left(  \frac{1}{\tau^{2}}-
\frac {1}{\tau_{m}^{2}}\right) \right]  \hat{\tau}^{0}\right. \nonumber\\
&  \left.  +\left[  \frac{1}{\tau}-\frac{\tau_{\mathrm{qp}}\tilde{\gamma}_{\mathbf{q}}}{2}\left(
\frac{1}{\tau^{2}}-\frac{1}{\tau_{m}^{2}}\right)  \right]  \hat{\tau}
^{3}\right. \nonumber\\
&  \left.  -\tau_{\mathrm{qp}}\tilde{\gamma}_{\mathbf{q}}\frac{\operatorname{Re}\eta_{+}}{\left\vert
\eta_{+}^{2}-1\right\vert }\left(  \frac{1}{\tau^{2}}-\frac{1}{\tau_{m}^{2}
}\right)  \hat{\tau}^{1}\right\}.
\end{align}

Similarly, in order to compute the block $\hat{C}^{\mathrm{(sm)}}_{B}$, we derive relation
\begin{align*}
2\nu\hat{u}^{B}\hat{S}^{B}  &  =\tau_{\mathrm{qp}}\tilde{\gamma}_{\mathbf{q}}\left[  \left(
\frac{1}{2\tau}+\frac{\chi_2}{2\tau_{m}}\right)  \hat{\tau}^{0}-\left(
\frac{\chi_2}{2\tau}+\frac{1}{2\tau_{m}}\right)  \hat{\tau}^{3}\right. \\
&  \left.  -\left(  \frac{1}{\tau}\hat{\tau}^{1}-\frac{1}{\tau_{m}}i\hat{\tau
}^{2}\right)  \frac{\operatorname{Im}\eta_{+}}{\left\vert \eta_{+}
^{2}-1\right\vert }\right]
\end{align*}
with $\chi_2=\left(  |\eta_{+}|^{2}-1\right)  /\left\vert
\eta_{+}^{2}-1\right\vert $. This allows us to present solution for
$\hat{C}^{\mathrm{(sm)}}_{B}$ in the form
\begin{align*}
&  \hat{C}^{\mathrm{(sm)}}_{B}=\left\{  \left[  1-\tau_{\mathrm{qp}}\tilde{\gamma}_{\mathbf{q}}\left(  \frac
{1}{2\tau}+\frac{\chi_2}{2\tau_{m}}\right)  \right]  \hat{\tau}^{0}\right. \\
&  \left.  +\tau_{\mathrm{qp}}\tilde{\gamma}_{\mathbf{q}}\left[  \left(  \frac{\chi_2}{2\tau
}\!+\!\frac{1}{2\tau_{m}}\right)  \hat{\tau}^{3}\!+\!\left(  \frac{1}{\tau
}\hat{\tau}^{1}\!-\!\frac{i}{\tau_{m}}\hat{\tau}^{2}\right)  \frac
{\operatorname{Im}\eta_{+}}{\left\vert \eta_{+}^{2}-1\right\vert }\right]
\right\}  ^{-1}\\
&  \times\frac{1}{2}\left(  -u\hat{\tau}^{3}+u_{m}\hat{\tau}^{0}\right)  .
\end{align*}
After straightforward algebra one can reduce it to
\begin{align}
&\hat{C}^{\mathrm{(sm)}}_{B}\!=\!\frac{1/4\pi\nu}{1\!-\!\tau_{\mathrm{qp}}\tilde{\gamma}_{\mathbf{q}}\left(
1/\tau\!+\!\chi_2/\tau_{m}\right)}\left\{\hat{\tau}^{0}\!\left[\frac{1}{\tau_{m}}\right.\right.\nonumber\\
&\left.\left.+\frac{\chi_2\tau_{\mathrm{qp}}\tilde{\gamma}_{\mathbf{q}}}{2}\!
\left(\frac{1}{\tau^{2}}\!-\!\frac{1}{\tau_{m}^{2}}\right) \right]
\!-\!\hat{\tau}^{3}\!\left[\frac{1}{\tau}\!-\!\frac{\tau_{\mathrm{qp}}
\tilde{\gamma}_{\mathbf{q}}}{2}\!\left(\frac{1}{\tau^{2}}\!-\!\frac{1}{\tau_{m}^{2}
}\right) \right]  \right. \nonumber\\
&  \left. +i\tau_{\mathrm{qp}}\tilde{\gamma}_{\mathbf{q}}\hat{\tau}^{2}\left(
\frac{1}{\tau^{2}}\!-\!\frac{1}{\tau_{m}^{2}}\right)  \frac{\operatorname{Im}
\eta_{+}}{\left\vert \eta_{+}^{2}\!-\!1\right\vert }\right\}  .\label{CB}
\end{align}

\subsection{Calculation of Hikami boxes \label{App-Hikami-sm}}

Let us calculate Hikami boxes (\ref{HikamiSP}) in the case of s-wave
superconductor with magnetic impurities, i.e. using the matrix Green's
function (\ref{GRAmp}). For the component $B_{000}^{\left(  \mathrm{sm}
\right)  }$ one finds\begin{widetext}
\begin{align*}
B_{000}^{\left( \mathrm{sm}\right) }  =-\nu\int d\xi\frac{\tilde{E}_{+}^{2}\tilde{E}_{-}}{\left[
\xi^{2}-\left(  \operatorname{Re}\sqrt{\tilde{E}_{+}^{2}-\tilde{\Delta}
_{+}^{2}}+\frac{i}{2\tau_{\mathrm{qp}}}\right)  ^{2}\right]  ^{2}\left[  \xi^{2}-\left(
\operatorname{Re}\sqrt{\tilde{E}_{+}^{2}-\tilde{\Delta}_{+}^{2}}-\frac
{i}{2\tau_{\mathrm{qp}}}\right)  ^{2}\right]  }
=-\frac{\nu\tilde{E}_{+}^{2}\tilde{E}_{-}}{\left[  \operatorname{Re}
\sqrt{\tilde{E}_{+}^{2}-\tilde{\Delta}_{+}^{2}}\right]  ^{5}}I^{\left(
0\right)  }
\end{align*}
\vspace{-0.1in}
\end{widetext}
with
\begin{align*}
I^{\left(  \mu\right)  }\left(  \kappa\right) & =\int_{-\infty}^{\infty}
\frac{x^{\mu}dx}{\left[  x^{2}-\left(  1+i\kappa\right)  ^{2}\right]
^{2}\left[  x^{2}-\left(  1-i\kappa\right)  ^{2}\right]  },\\
\kappa&=\frac{1}{2\tau_{\mathrm{qp}}\operatorname{Re}\sqrt{\tilde{E}_{+}^{2}
-\tilde{\Delta}_{+}^{2}}}.
\end{align*}
This integral can be easily evaluated for relevant values of $\mu$:
\[
I^{\left(  \mu\right)  }=\frac{i\pi}{8\kappa^{2}(1+i\kappa)}
\left\{
\begin{tabular}
[c]{l}
$\frac{(1+3i\kappa)}{(1+i\kappa)(1+\kappa^{2})}$, for $\mu=0$\\
$1$,  for $\mu=2$
\end{tabular}
\right.  .
\]
In result
\[
B_{000}^{\left(  \mathrm{sm}\right)  }=-\frac{i\pi\nu\tau_{\mathrm{qp}}^{2}}{2}
\frac{\eta_{+}^{2}\eta_{-}\left(  \sqrt{\eta_{+}^{2}-1}+\frac{i}{\tau
_{\mathrm{qp}}\tilde{\Delta}_{+}}\right)  }{\left(  \eta_{+}^{2}-1\right)  ^{3/2}
\sqrt{\eta_{-}^{2}-1}}.
\]

For the convenient representation of the remaining components $B_{imj}$ we
introduce notation $\zeta_{+}=\tilde{\Delta}_{+}/\tilde{E}_{+}=1/\eta_{+}$.
When the subscripts are set to $k,l,m=0,1$ one finds
\[
B_{kml}^{\left(  \mathrm{sm}\right)  }=-\frac{i\pi\nu\tau_{\mathrm{qp}}^{2}}{2}
\frac{\tilde{E}_{+}^{2}\tilde{E}_{-}\left(  \sqrt{\tilde{E}_{+}^{2}
-\tilde{\Delta}_{+}^{2}}+\frac{i}{\tau_{\mathrm{qp}}}\right)  }{\left(  \tilde{E}
_{+}^{2}-\tilde{\Delta}_{+}^{2}\right)  ^{3/2}\sqrt{\tilde{E}_{-}^{2}
-\tilde{\Delta}_{-}^{2}}}\zeta_{+}^{k+l}\zeta_{-}^{m}.
\]
Other non-zero components are
\begin{widetext}
\vspace{-0.2in}
\begin{align*}
\left(
\begin{tabular}
[c]{l}
$B_{033}^{\left( \mathrm{sm}\right) }$\\
$B_{303}^{\left( \mathrm{sm}\right) }$
\end{tabular}
\ \right)   &  =-\nu\int d\xi\frac{\tilde{E}_{\pm}\xi^{2}}{\left[  \xi
^{2}-\left(  \operatorname{Re}\sqrt{\tilde{E}_{+}^{2}-\tilde{\Delta}_{+}^{2}
}+\frac{i}{2\tau_{\mathrm{qp}}}\right)  ^{2}\right]  ^{2}\left[  \xi^{2}-\left(
\operatorname{Re}\sqrt{E^{2}-\Delta^{2}}-\frac{i}{2\tau_{\mathrm{qp}}}\right)
^{2}\right]  }\\
&  =-\frac{\nu\tilde{E}_{\pm}}{\left(  \operatorname{Re}\sqrt{\tilde{E}
_{+}^{2}-\tilde{\Delta}_{+}^{2}}\right)  ^{3}}I^{(2)}
=-\frac{i\pi\nu\tau_{\mathrm{qp}}^{2}}{2}\frac{\tilde{E}_{\pm}}{\sqrt{\tilde{E}_{+}
^{2}-\tilde{\Delta}_{+}^{2}}}
\end{align*}
\vspace{-0.15in}
\end{widetext}
and
\[
B_{i33}^{\left(  \mathrm{sm}\right)  }=\zeta_{+}^{i}B_{033}^{\left(
\mathrm{sm}\right)  }\text{,\ }B_{3i3}^{\left(  \mathrm{sm}\right)  }
=\zeta_{-}^{i}B_{303}^{\left(  \mathrm{sm}\right)  }\text{ for }i=0,1.
\]

\subsection{Quasiparticle relaxation time $\tau_{\mathrm{qp}}$ \label{App-QPtime}}

The quasiparticle relaxation time $1/\tau_{\mathrm{qp}}(E)=2\operatorname{Im}\sqrt
{\tilde{E}_{+}^{2}-\tilde{\Delta}_{+}^{2}}$ determines the value of diffusion
constant of quasiparticles in presence of magnetic impurities. In this Appendix
we derive a useful formula for this parameter. From the definition of
$\tilde{\Delta}_{+}$, Eq.\ (\ref{DRAmp}), one obtains the relation
\begin{align}
1/\tau_{\mathrm{qp}}(E)  &  =2\operatorname{Im}\left[  \tilde{\Delta}^{R}\sqrt{\eta
_{+}^{2}-1}\right] \nonumber\\
&  =2\Delta\operatorname{Im}\sqrt{\eta_{+}^{2}-1}+\frac{1}{\tau}-\frac{1}
{\tau_{m}}.\label{RelTimeInit}
\end{align}
Presenting the parameter $\eta_{+}$ as the sum of its real and imaginary
parts, $\eta_{+}=\eta_{r}+i\eta_{i}$, we find Eq.(\ref{SelfConsuRA}) from the
following relation for the imaginary part
\begin{align*}
\eta_{i}  &  =\frac{1}{\tau_{m}\Delta}\operatorname{Re}\frac{\eta_{+}}
{\sqrt{\eta_{+}^{2}-1}}\\
&  =\frac{1}{\tau_{m}\Delta}\frac{\eta_{r}\operatorname{Re}\left[  \sqrt
{\eta_{+}^{2}-1}\right]  +\eta_{i}\operatorname{Im}\left[  \sqrt{\eta_{+}
^{2}-1}\right]  }{|\eta_{+}^{2}-1|}.
\end{align*}
Using this result, the factor $\operatorname{Im}\sqrt{\eta_{+}^{2}-1}$ in the
right-hand side of Eq.\ (\ref{RelTimeInit}) can be transformed as,
\begin{align*}
\operatorname{Im}\sqrt{\eta_{+}^{2}\!-\!1}  &  \!=\!\frac{\eta_{r}\eta_{i}
}{\operatorname{Re}\sqrt{\eta_{+}^{2}\!-\!1}}=\frac{1}{\tau_{m}\Delta}\frac
{\eta_{r}^{2}\!+\!\left(\operatorname{Im}\left[\sqrt{\eta_{+}^{2}\!-\!1}\right]
\right)  ^{2}}{|\eta_{+}^{2}\!-\!1|}\\
&  =\frac{1}{2\tau_{m}\Delta}\left(  1+\frac{|\eta_{+}|^{2}+1}{|\eta_{+}
^{2}-1|}\right)  .
\end{align*}
Substituting this result into Eq.\ (\ref{RelTimeInit}), we finally obtain
presentation
\begin{align}
1/\tau_{\mathrm{qp}}(E)=\frac{1}{\tau}+\frac{1}{\tau_{m}}\frac{|\eta_{+}|^{2}+1}
{|\eta_{+}^{2}-1|}, \label{tauqp}
\end{align}
which was used in Appendix \ref{App-Cooperon-sm}. %
\vspace{-0.15in}

\section{Trace of five Pauli matrices \label{App-trace-Pauli}}

First of all let us symmetrize trace (\ref{VertexSC}) with respect to two
indices:
\[
\frac{1}{2}\mathrm{Tr}\left(  \hat{\tau}^{i}\hat{\tau}^{k}\hat{\tau}^{n}
\hat{\tau}^{m}\hat{\tau}^{j}\right)  =\frac{1}{4}\mathrm{Tr}[\left(  \hat
{\tau}^{j}\hat{\tau}^{i}+\hat{\tau}^{i}\hat{\tau}^{j}\right)  \hat{\tau}
^{k}\hat{\tau}^{n}\hat{\tau}^{m}].
\]
Let us recall that
\begin{align*}
\hat{\tau}^{i}\hat{\tau}^{k}  &  =\delta_{i0}\hat{\tau}^{k}+\delta_{k0}
\hat{\tau}^{i}-\delta_{i0}\delta_{k0}\hat{\tau}^{0}\\
&  +\left(  1-\delta_{i0}\right)  \left(  1-\delta_{k0}\right)  \left[
\delta_{ik}\hat{\tau}^{0}+i\sum_{s=1}^{3}\varepsilon_{iks}\hat{\tau}
^{s}\right] \\
\hat{\tau}^{i}\hat{\tau}^{j}\!+\!\hat{\tau}^{j}\hat{\tau}^{i}  &  =\! 2\delta
_{ij}\hat{\tau}^{0}\!+\!2\delta_{i0}\left( \hat{\tau}^{j}\!-\!\delta_{j0}\hat{\tau
}^{0}\right) \!+\!2\delta_{j0}\left(  \hat{\tau}^{i}\!-\!\delta_{i0}\hat{\tau}
^{0}\right)
\end{align*}
and write down the trace of three Pauli matrices:
\begin{align*}
\frac{1}{2}\mathrm{Tr}[\hat{\tau}^{k}\hat{\tau}^{n}\hat{\tau}^{m}]  &
=\delta_{m0}\delta_{kn}+\delta_{k0}\delta_{mn}+\delta_{n0}\delta_{mk}\\
&  -2\delta_{m0}\delta_{k0}\delta_{n0}+i\varepsilon_{0knm}
\end{align*}
($\varepsilon_{iks}$ and $\varepsilon_{0knm}$ here are 3D and 4D Levi-Civita symbols).

A further step is the calculation of the trace of four Pauli matrices:
\begin{align*}
&  \frac{1}{2}\mathrm{Tr}[\left(  \hat{\tau}^{j}-\delta_{j0}\hat{\tau}
^{0}\right)  \hat{\tau}^{k}\hat{\tau}^{n}\hat{\tau}^{m}]=\frac{1}
{2}\mathrm{Tr}[\left(  \delta_{jk}-\delta_{j0}\delta_{k0}\right)  \hat{\tau
}^{n}\hat{\tau}^{m}\\
&  +\delta_{k0}\left(  \hat{\tau}^{j}-\delta_{j0}\hat{\tau}^{0}\right)
\hat{\tau}^{n}\hat{\tau}^{m}+i\sum_{s=0}^{3}\varepsilon_{0jks}\hat{\tau}
^{s}\hat{\tau}^{n}\hat{\tau}^{m}]\\
&  =\left(  1-\delta_{j0}\right)  \left[  \delta_{jk}\delta_{nm}+\delta
_{jm}\delta_{kn}-\delta_{jn}\left(  \delta_{km}-2\delta_{k0}\delta
_{m0}\right)  \right. \\
&  \left.  +i\varepsilon_{0jnm}\delta_{k0}+i\varepsilon_{0jkn}\delta
_{m0}+i\varepsilon_{0jkm}\delta_{n0}\right]  .
\end{align*}
We used the relation
\begin{align*}
&  \left(  \hat{\tau}^{j}-\delta_{j0}\hat{\tau}^{0}\right)  \hat{\tau}^{n}\\
&  =\left(  \delta_{jn}-\delta_{j0}\delta_{n0}\right)  \hat{\tau}^{0}
+\delta_{n0}\left(  \hat{\tau}^{j}-\delta_{j0}\hat{\tau}^{0}\right)
+i\sum_{s=0}^{3}\varepsilon_{0jns}\hat{\tau}^{s}
\end{align*}
and the expression for product of Levi-Civita symbols
\[
\sum_{s=0}^{3}\varepsilon_{0jks}\varepsilon_{0snm}=(1-\delta_{j0}
)(1-\delta_{k0})\left(  \delta_{jn}\delta_{km}-\delta_{jm}\delta_{kn}\right)
.
\]
Finally one finds the required symmetrized trace of five Pauli matrices:
\begin{align*}
&  \frac{1}{4}\mathrm{Tr}[\hat{\tau}^{i}\hat{\tau}^{k}\hat{\tau}^{n}\hat{\tau
}^{m}\hat{\tau}^{j}+\hat{\tau}^{j}\hat{\tau}^{k}\hat{\tau}^{n}\hat{\tau}
^{m}\hat{\tau}^{i}]=\\
&  =\delta_{ij}\left(  \delta_{m0}\delta_{kn}+\delta_{k0}\delta_{mn}
+\delta_{n0}\delta_{mk}-2\delta_{m0}\delta_{k0}\delta_{n0}+i\varepsilon
_{0knm}\right) \\
&  +\delta_{i0}\left(  1-\delta_{j0}\right)  \left[  \delta_{jk}\delta
_{nm}+\delta_{jm}\delta_{kn}-\delta_{jn}\left(  \delta_{km}-2\delta_{k0}
\delta_{m0}\right)  \right. \\
&  +i\varepsilon_{0jnm}\delta_{k0}+i\varepsilon_{0jkn}\delta_{m0}
+i\varepsilon_{0jkm}\delta_{n0}\\
&  +\delta_{j0}\left(  1-\delta_{i0}\right)  \delta_{ik}\delta_{nm}
+\delta_{im}\delta_{kn}-\delta_{in}\left(  \delta_{km}-2\delta_{k0}\delta
_{m0}\right) \\
&  \left.  +i\varepsilon_{0inm}\delta_{k0}+i\varepsilon_{0ikn}\delta
_{m0}+i\varepsilon_{0ikm}\delta_{n0}\right]  .
\end{align*}

\section{d-wave superconductors \label{App-dwave}}

\subsection{Calculation of d-wave cooperon \label{App-Cooperon-d}}

The cooperon corresponding to propagation of quasi-particles for
superconductors with d-wave symmetry of the order parameters
can be found from the same Eq.\ (\ref{EqSCcoop}) as it
was done above. We again employ its expansion in terms of Pauli matrices (see
Eq.\ (\ref{CooperonPauli})). The analogous expansion $\Pi_{\nu\nu_{1}
}^{\gamma\gamma_{1}}=\Pi_{mn}\tau_{\nu\nu_{1}}^{m}\tau_{\gamma\gamma_{1}}^{n}
$ we apply also to the polarization function, the only nonzero components of
which are $\Pi_{00}$, $\Pi_{11}$, and $\Pi_{33}$.

According to Refs. \onlinecite{YashenkinPRL01,YangPRB04} only the diagonal
components of such cooperon are singular. Keeping only these components, we
find that they obey relatively simple linear system of equations:
\begin{subequations}
\begin{align}
(1-u\Pi_{33})C^{\mathrm{(sd)}}_{00}-u\Pi_{00}C^{\mathrm{(sd)}}_{33}+u\Pi_{11}C^{\mathrm{(sd)}}_{22}  &  =0\label{Cd00}\\
(1-u\Pi_{33})C^{\mathrm{(sd)}}_{33}-u\Pi_{00}C^{\mathrm{(sd)}}_{00}-u\Pi_{11}C^{\mathrm{(sd)}}_{11}  &  =u\label{Cd33}\\
(1-u\Pi_{33})C^{\mathrm{(sd)}}_{11}+u\Pi_{00}C^{\mathrm{(sd)}}_{22}-u\Pi_{11}C^{\mathrm{(sd)}}_{33}  &  =0\label{Cd11}\\
(1-u\Pi_{33})C^{\mathrm{(sd)}}_{22}+u\Pi_{00}C^{\mathrm{(sd)}}_{11}+u\Pi_{11}C^{\mathrm{(sd)}}_{00}  &  =0.\label{Cd22}
\end{align}
In accordance with Refs.\ \onlinecite{YashenkinPRL01,YangPRB04}, the ansatz
\end{subequations}
\begin{equation}
C^{\mathrm{(sd)}}_{00}=C^{\mathrm{(sd)}}_{11}=-C^{\mathrm{(sd)}}_{22}=C^{\mathrm{(sd)}}_{33}\equiv C_{\mathrm{d}}\label{SingCdRel}
\end{equation}
reduces the right hand side of all four equations to $(1-u\Pi_{ii})C_{\mathrm{d}}$. As
we will see below, $u\Pi_{ii}(q,\Omega) \rightarrow1$ for $q,\Omega
\rightarrow0$ meaning that the relations (\ref{SingCdRel}) correspond to the
singular eigenvector. Taking the linear combination of equations
(\ref{Cd00})+(\ref{Cd33})+(\ref{Cd11})-(\ref{Cd22}), we obtain
\[
(1-u\Pi_{ii})\left(  C^{\mathrm{(sd)}}_{00}+C^{\mathrm{(sd)}}_{33}+C^{\mathrm{(sd)}}_{11}-C^{\mathrm{(sd)}}_{22}\right)  =u
\]
which gives
\begin{equation}
C_{\mathrm{d}}(q)=\left(  u/4\right)  /\left[  1-u\Pi_{ii}(q,0)\right] .\label{SingCd}
\end{equation}
We noticed, however, that this result is 4 times smaller than the one reported
in Refs.\ \onlinecite{YashenkinPRL01,YangPRB04}. A detailed presentation of
derivation given in Ref.\ \onlinecite{YangPRB04} allows us to trace the origin
of this discrepancy. The authors of Ref.\ \onlinecite{YangPRB04} obtained their
expression for $C_{\mathrm{d}}$ by substitution of the relations
(\ref{SingCdRel}) between the singular parts of $C^{\mathrm{(sd)}}_{ii}$ into
Eq.\ (\ref{Cd33}), what led to $C_{\mathrm{d}}=u/\left[
1-u\Pi_{ii}(q,0)\right]  $. This step, however, is problematic since side by
side with the singular parts, $C^{\mathrm{(sd)}}_{ii}$ also contain regular
contributions: $C^{\mathrm{(sd)}}_{ii}=\alpha_{i}C^{\mathrm{(sd)}}_{d}+c_{i}$,
with $\alpha_{2}=-1$; $\alpha_{i}=1$ for $i\neq2$ and $c_{i}$ are constants.
Substituting such presentation into Eq.\ (\ref{Cd33}), we immediately see that
the constants $c_{i}$ contribute to the nominator of $C_{\mathrm{d}}$. As
follows from the result (\ref{SingCd}), this contribution amounts to its four
times reduction.

Let us pass to evaluation of the trace of the polarization function:
\begin{widetext}
\begin{equation}
\Pi_{ii}(q,\Omega)=\int\frac{d^{2}\mathbf{p}}{(2\pi)^{2}}\frac
{\tilde{\epsilon}^{2}+(\gamma-i\tilde{\Omega}/2)^{2}+\varepsilon_{\frac{\mathbf{q}}{2}-\mathbf{p}
}\varepsilon_{\frac{\mathbf{q}}{2}+\mathbf{p}}}{\left[  \left(  \tilde{\epsilon
}+\tilde{\Omega}/2+i\gamma\right)  ^{2}-\varepsilon_{\frac{\mathbf{q}}{2}-\mathbf{p}}^{2}\right]
\left[  \left(  \tilde{\epsilon}-\tilde{\Omega}/2-i\gamma\right)  ^{2}-\varepsilon_{\frac
{\mathbf{q}}{2}+\mathbf{p}}^{2}\right]  }
\label{TracePolOperDwave}
\end{equation}
\vspace{-0.1in}
\end{widetext}
with $\tilde{\Omega}=d\tilde{\epsilon}/dE\ \Omega$. Note that
the vanishing of $1-u\Pi_{ii}(q,\Omega)$ for $q,\Omega\rightarrow0$ leading to
diffusive behavior follows from general identity
\begin{equation}
u\Pi_{ii}(0,0)=-\frac{\Sigma_{0}^{+}-\Sigma_{0}^{-}}{2i\gamma}
=1,\label{condition}
\end{equation}
which can be derived from the above definition of $\Pi_{ii}(q,\Omega)$.

The next step is to perform the expansion of the polarization operator over
$q$ up to quadratic term:
\begin{equation}
u\Pi_{ii}(q,0)-1=-\frac{\langle v^{2}\rangle}{4\gamma^{2}}\frac{1+\frac
{\tilde{\epsilon}}{\gamma}\arctan\frac{\tilde{\epsilon}}{\gamma}}{\ln
\frac{\Delta_{0}}{\sqrt{\tilde{\epsilon}^{2}+\gamma^{2}}}+\frac{\tilde
{\epsilon}}{\gamma}\arctan\frac{\tilde{\epsilon}}{\gamma}}q^{2}
.\label{PolOpExpan}
\end{equation}
with $\langle v^{2}\rangle\equiv(v_{\mathrm{f}}^{2}+v_{\mathrm{g}}^{2})/2$.
For the trace of the polarization operator at finite frequency and $q=0$,
$\Pi_{ii}(0,\Omega)$, we derive the following relation
\[
u\Pi_{ii}(0,\Omega)=-\frac{\Sigma_{0}^{+}(E+\frac{\Omega}{2})-\Sigma_{0}
^{-}(E-\frac{\Omega}{2})}{\tilde{\Omega}+2i\gamma}.
\]
As $\mathrm{Re}\Sigma_{0}^{\pm}=E-\tilde{\epsilon}$, the difference
$\Sigma_{0}^{+}(E+\frac{\Omega}{2})-\Sigma_{0}^{-}(E-\frac{\Omega}{2})$ can be
represented as
\begin{align*}
\Sigma_{0}^{+}\left(  E\!+\!\frac{\Omega}{2}\right)  -\Sigma_{0}^{-}\left(
E\!-\!\frac{\Omega}{2}\right)   &  \approx\Omega\left(  1\!-\!\frac
{d\tilde{\epsilon}}{dE}\right)  -2i\gamma\\
&  =\Omega-\tilde{\Omega}-2i\gamma,
\end{align*}
leading to a very simple result for small $\Omega$
\[
\Pi_{ii}(0,\Omega)\approx1+i\Omega/2\gamma.
\]
Collecting terms, we obtain
\begin{equation}
u\Pi_{ii}(q,\Omega)=1+\frac{1}{2\gamma}\left(  i\Omega-\mathcal{D}q^{2}\right)
\label{PolOpTraceDwave}
\end{equation}
with energy-dependent diffusion coefficient
\begin{equation}
\mathcal{D}\left(  E\right)  =\frac{\langle v^{2}\rangle}{2\gamma}\frac{1+\frac
{\tilde{\epsilon}}{\gamma}\arctan\frac{\tilde{\epsilon}}{\gamma}}{\ln
\frac{\Delta_{0}}{\sqrt{\tilde{\epsilon}^{2}+\gamma^{2}}}+\frac{\tilde
{\epsilon}}{\gamma}\arctan\frac{\tilde{\epsilon}}{\gamma}}.
\end{equation}
Its value for zero-energy was first obtained in
Refs.\ \onlinecite{YashenkinPRL01,YangPRB04}. Substituting
Eq.\ (\ref{PolOpTraceDwave}) into Eq. (\ref{SingCd}) and using relation
$\gamma= (\pi/2) u N_{ex,d}$ valid for all energies, we obtain the presentation of
the cooperon in Eq.\ (\ref{Coop-d1}) of the main text.

\subsection{Calculation of d-wave Hikami boxes \label{App-Hikami-d}}

In this appendix we calculate the Hikami boxes for two-dimensional superconductor with d-wave symmetry of the
order parameter:
\begin{align*}
\mathcal{B}_{imj}^{\left(  \mathrm{d}\right)  } &  =\int d\mathbf{R}_{1}\int
d\mathbf{R}_{2}G_{i}^{R}(\mathbf{R}_{1})G_{m}^{A}(\mathbf{R}_{1}
+\mathbf{R}_{2})G_{j}^{R}(\mathbf{R}_{2})\\
&  =\int\frac{d^2\mathbf{k}}{(2\pi)^2}  G_{i}^{R}(\mathbf{k})G_{m}^{A}
(\mathbf{k})G_{j}^{R}(\mathbf{k}).
\end{align*}
The Green's function for d-wave superconductor is determined by Eq.\ (\ref{gd}), which we rewrite in the form
\[
G_{k}^{R,A}\left(  \epsilon\right)  =\frac{g_{i}^{R,A}\hat{\tau}^{i}}{\left[
\tilde{\epsilon}\pm i\gamma\right]  ^{2}-\varepsilon_{k}^{2}}
\]
with
\[
\left(
\begin{array}
[c]{c}
g_{0}^{R,A}\\
g_{1}^{R,A}\\
g_{2}^{R,A}\\
g_{3}^{R,A}
\end{array}
\right)  =\left(
\begin{array}
[c]{c}
\tilde{\epsilon}\pm i\gamma\\
\Delta_{k}=v_{\mathrm{g}}k_{y}\\
0\\
\xi_{k}=v_{\mathrm{f}}k_{x}
\end{array}
\right)  .
\]
Hence the block $\mathcal{B}_{imj}^{\left(  \mathrm{d}\right)  }$ can be
rewritten as
\begin{widetext}
\vspace{-0.2in}
\begin{equation*}
\mathcal{B}_{imj}^{\left( \mathrm{d}\right) }
=\frac{N_{n}}{v_{\mathrm{f}}v_{\mathrm{g}}}\int \int \frac{
d\Delta _{k}}{2\pi }\frac{d\xi _{k}}{2\pi }\frac{g_{i}^{R}( \Delta _{k},\xi _{k})\
g_{m}^{A}( \Delta _{k},\xi _{k})\
g_{j}^{R}( \Delta _{k},\xi _{k}) }{\left[ \alpha ^{2}-\left(
\Delta _{k}^{2}+\xi _{k}^{2}\right) \right] ^{2}\left[ (\alpha ^{\ast
})^{2}-\left( \Delta _{k}^{2}+\xi _{k}^{2}\right) \right] }
\end{equation*}
\vspace{-0.15in}
\end{widetext}
with $\alpha=\tilde{\epsilon}+i\gamma$. The only non-zero
components are $\mathcal{B}_{000}^{\left(  \mathrm{d}\right)  },\mathcal{B}
_{011}^{\left(  \mathrm{d}\right)  },\mathcal{B}_{101}^{\left(  \mathrm{d}
\right)  },\mathcal{B}_{110}^{\left(  \mathrm{d}\right)  },\mathcal{B}
_{033}^{\left(  \mathrm{d}\right)  },\mathcal{B}_{303}^{\left(  \mathrm{d}
\right)  },\mathcal{B}_{330}^{\left(  \mathrm{d}\right)  }$ and to get the
explicit expressions for them one has to carry out the integrals:
\[
J^{(\mu)}\!=\!\int\int\frac{d\Delta_{k}}{2\pi}\frac{d\xi_{k}}{2\pi}\frac
{\xi_{k}^{\mu}}{\left[  \alpha^{2}\!-\!\left(  \Delta_{k}^{2}\!+\xi_{k}
^{2}\right)  \right]  ^{2}\left[  (\alpha^{\ast})^{2}\!-\!\left(  \Delta
_{k}^{2}\!+\xi_{k}^{2}\right)  \right]  }
\]
with $\mu=0,2$. Then the blocks $B_{imj}$ are expressed as
\begin{align*}
\mathcal{B}_{000}^{\left(  \mathrm{d}\right)  } &  =N_{n}\frac{\alpha^{2}
\alpha^{\ast}}{v_{\mathrm{f}}v_{\mathrm{g}}}J^{\left(  0\right)  },\\
\mathcal{B}_{101}^{\left(  \mathrm{d}\right)  } &  =\mathcal{B}_{303}^{\left(
\mathrm{d}\right)  }=N_{n}\frac{\alpha^{\ast}}{v_{\mathrm{f}}v_{\mathrm{g}}}J^{(2)},\\
\mathcal{B}_{330}^{\left(  \mathrm{d}\right)  } &  =\mathcal{B}_{033}^{\left(
\mathrm{d}\right)  }=\mathcal{B}_{011}^{\left( \mathrm{d}\right)
}=\mathcal{B}_{110}^{\left(  \mathrm{d}\right)  }=N_{n}\frac{\alpha}{v_{\mathrm{f}}v_{\mathrm{g}}
}J^{\left(  2\right)  }.
\end{align*}
The integrals $J^{\left(  \mu\right)  }$ can be computed explicitly as
\[
J^{(0)}=-\frac{i}{16\pi\left(  \tilde{\epsilon}+i\gamma\right)  ^{2}
\tilde{\epsilon}\gamma}\left[  1+\frac{\left(  \tilde{\epsilon}+i\gamma
\right)  ^{2}}{\tilde{\epsilon}\gamma}\arctan\frac{\tilde{\epsilon}}{\gamma
}\right]
\]
and
\begin{align*}
J^{(2)} &  =-\frac{i}{32\pi\tilde{\epsilon}\gamma}\left[1\!+\!\frac{\left(
\tilde{\epsilon}-i\gamma\right)^{2}}{\tilde{\epsilon}\gamma}\arctan
\frac{\tilde{\epsilon}}{\gamma}\right] \\
& =-\left[  J^{\left(  0\right)  }\right]^{\ast}\frac{\left[
\tilde{\epsilon}-i\gamma\right]  ^{2}}{2}.
\end{align*}

\end{document}